\documentclass[a4paper,12pt]{article}
\usepackage{expl3}
\usepackage{amsmath,amssymb,mathrsfs,amsthm,tikz,shuffle,paralist}
\usepackage{dsfont}
\usepackage{amsbsy}

\usepackage{xcolor}
\usepackage{relsize}
\usepackage{subfig}

\usepackage[font={small},labelfont=bf]{caption}

\usepackage[compat=1.1.0]{tikz-feynman}
\usetikzlibrary{shapes.misc}
\usetikzlibrary{cd}
\usepackage{todonotes}
\usepackage{stackengine}
\usepackage[bottom]{footmisc}
\usepackage{booktabs}

\usetikzlibrary{arrows}
\newcommand{\midarrow}{\tikz \draw [-{Stealth[length=2.2mm]}](0,0) -- (.1,0);}

\newcommand{\nint}{{n_{\text{int}}}}

\newcommand{\nISP}{{n_{\text{ISP}}}}
\newcommand{\LS}{{\text{LS}}}

\newcommand{\diagramnumberingsign}{}

\usepackage{array}
\newcommand{\PreserveBackslash}[1]{\let\temp=\\#1\let\\=\temp}
\newcolumntype{C}[1]{>{\PreserveBackslash\centering}p{#1}}
\newcolumntype{R}[1]{>{\PreserveBackslash\raggedleft}p{#1}}
\newcolumntype{L}[1]{>{\PreserveBackslash\raggedright}p{#1}}

\usepackage{url}

\usepackage{breakurl}
\usepackage[colorlinks=true
,urlcolor=blue
,anchorcolor=blue
,citecolor=blue
,filecolor=blue
,linkcolor=blue
,menucolor=blue
,pagecolor=blue
,linktocpage=true
,pdfproducer=medialab
,pdfa=true
]{hyperref}
\hypersetup{breaklinks=true}

\usepackage[utf8]{inputenc}
\usepackage{graphicx}

\usepackage{jheppub}

\begin{document}

\title{Classification of Feynman integral geometries for black-hole scattering at 5PM order}

\author[a]{Daniel Brammer,}
\emailAdd{mhr544@alumni.ku.dk}
\author[b,a]{Hjalte Frellesvig,}
\emailAdd{0025056@zju.edu.cn}
\author[a]{Roger Morales,}
\emailAdd{roger.morales@nbi.ku.dk}
\author[c,a]{and Matthias Wilhelm}
\emailAdd{mwilhelm@imada.sdu.dk}

\affiliation[a]{%
Niels Bohr International Academy, Niels Bohr Institute, Copenhagen University, 2100 Copenhagen \O{}, Denmark}
\affiliation[b]{%
Zhejiang Institute of Modern Physics, School of Physics, Zhejiang University, Hangzhou 310027, China}
\affiliation[c]{%
Center for Quantum Mathematics, Department of Mathematics and Computer Science, University of Southern Denmark, 5230 Odense M, Denmark}

\date{\today}%

\abstract{%
We provide a complete classification of the Feynman integral geometries relevant to the scattering of two black holes at fifth order in the post-Minkowskian (PM) expansion, i.e.\ at four loops. The analysis includes integrals relevant to both the conservative and dissipative dynamics, as well as to all orders in the self-force (SF) expansion, i.e.\ the 0SF, 1SF and 2SF orders. By relating the geometries of integrals across different loop orders and integral families, we find that out of the 16,596 potentially contributing integral topologies, only 70 need to be analyzed in detail. By further computing their leading singularities using the loop-by-loop Baikov representation, we show that there only appear two different three-dimensional Calabi--Yau geometries and two different K3 surfaces at this loop order, which together characterize the space of functions beyond polylogarithms to which the 5PM integrals evaluate.
}

\maketitle

\section{Introduction}
\label{sec:intro}

The upcoming third-generation gravitational-wave observatories such as the Einstein Telescope~\cite{Punturo:2010zz,Abac:2025saz} and Cosmic Explorer~\cite{Reitze:2019iox,Evans:2021gyd} will provide a wealth of high-precision experimental data for gravitational waves from binary systems of black holes and neutron stars, requiring high-precision theoretical predictions for its interpretation. A number of theoretical methods have been developed to calculate such predictions, ranging from numeric relativity~\cite{Pretorius:2005gq,Campanelli:2005dd} to analytical methods such as the effective one-body (EOB) approach~\cite{Buonanno:1998gg,Buonanno:2000ef} and the post-Newtonian (PN)~\cite{Goldberger:2004jt,Blanchet:2013haa,Levi:2018nxp}, post-Minkowskian (PM)~\cite{Damour:2016gwp,Buonanno:2022pgc} and self-force (SF)~\cite{Mino:1996nk,Quinn:1996am,Poisson:2011nh,Barack:2018yvs} expansions.

The PM approximation is an expansion in Newton's constant $G$ and it has been shown to be amenable to techniques from Quantum Field Theory (QFT), particularly from scattering amplitudes; see e.g.~refs.~\cite{Bjerrum-Bohr:2022blt,Buonanno:2022pgc} for a review. Specifically, one can calculate the corrections to the classical two-body dynamics at $n$PM order in terms of Feynman integrals relevant to the $2 \to 2$ scattering of black holes at $L=n-1$ loops~\cite{Damour:2016gwp,Bern:2019nnu,Bern:2019crd,Kalin:2020mvi,Mogull:2020sak}. In particular, analytic continuation~\cite{Kalin:2019rwq,Kalin:2019inp,Cho:2021arx,Dlapa:2024cje} and the EOB formalism~\cite{Damour:2016gwp} relate the bound problem to the scattering process and vice versa. Furthermore, this perturbative calculation can be naturally sorted into conservative and dissipative contributions~\cite{Damour:2020tta,DiVecchia:2021ndb,DiVecchia:2021bdo,Herrmann:2021tct,Bern:2021yeh,Dlapa:2022lmu,Jakobsen:2023hig,Driesse:2024feo}, as well as into different orders of the self-force expansion~\cite{Mino:1996nk,Quinn:1996am,Poisson:2011nh,Barack:2018yvs}, which is the expansion in the ratio of the black-hole masses.

Most Feynman integrals that have been computed to date can be expressed in terms of multiple polylogarithms~\cite{Chen:1977oja,Goncharov:1995ifj}, which are iterated integrals over the Riemann sphere generalizing classical polylogarithms. However, it is by now well established that multiple polylogarithms do not cover the entire class of functions appearing in Feynman integrals: also integrals over more intricate geometries appear; see ref.~\cite{Bourjaily:2022bwx} for a recent review. The first cases beyond the polylogarithmic class involve iterated integrals over elliptic kernels, which generalize the classical elliptic integrals, with paradigmatic examples being the massive sunrise~\cite{Sabry:1962rge,Broadhurst:1993mw,Laporta:2004rb,Adams:2013nia,Bloch:2013tra,Adams:2014vja} and the elliptic double-box integrals~\cite{Caron-Huot:2012awx,Bourjaily:2017bsb,Kristensson:2021ani,Morales:2022csr}. Beyond the elliptic case, there also appear integrals involving hyperelliptic curves~\cite{Huang:2013kh,Marzucca:2023gto,Duhr:2024uid}, as well as K3 surfaces and higher-dimensional Calabi--Yau (CY) manifolds~\cite{Brown:2010bw,Bourjaily:2018ycu,Bourjaily:2018yfy,Bonisch:2021yfw,Duhr:2022pch,Lairez:2022zkj,Pogel:2022vat,Duhr:2022dxb,Cao:2023tpx,McLeod:2023doa,Duhr:2023eld,Duhr:2024hjf,Duhr:2025ppd,Duhr:2025lbz,Maggio:2025jel}. Notably, examples of Feynman integrals depending on such non-trivial geometries can be found in precision calculations relevant for Standard Model particle phenomenology~\cite{Adams:2018bsn,Adams:2018kez,Broedel:2019hyg,Abreu:2019fgk,Duhr:2024bzt,Forner:2024ojj,Marzucca:2025eak,Becchetti:2025rrz,Bargiela:2025nqc} and for the black-hole scattering within the PM expansion of classical gravity~\cite{Bern:2021dqo,Dlapa:2021npj,Ruf:2021egk,Dlapa:2022wdu,Jakobsen:2023ndj,Frellesvig:2023bbf,Klemm:2024wtd,Frellesvig:2024zph,Driesse:2024feo,Frellesvig:2024rea}, which we will further explore in this paper.

Up to the third order in the PM expansion, corresponding to two-loop Feynman integrals, the black-hole scattering amplitude is expressible solely in terms of multiple polylogarithms~\cite{Bern:2019nnu,Bern:2019crd,Kalin:2020mvi,Mogull:2020sak,Herrmann:2021tct,DiVecchia:2021bdo,Bern:2021yeh,Dlapa:2022lmu,Jakobsen:2023hig}. At 4PM order, however, products of complete elliptic integrals appear in the conservative sector~\cite{Bern:2021dqo,Dlapa:2021npj,Jakobsen:2023ndj}, reflecting the emergence of a K3 surface in some of the contributing integrals~\cite{Ruf:2021egk,Dlapa:2022wdu,Frellesvig:2024zph}. 
In refs.~\cite{Frellesvig:2024zph,Frellesvig:2023bbf}, some of the present authors proposed a method to classify the geometries -- and thus the corresponding special functions -- that occur at higher orders in the PM expansion, in particular identifying a three-dimensional Calabi--Yau geometry at 5PM 2SF order in the conservative sector~\cite{Frellesvig:2023bbf,Frellesvig:2024rea}. Subsequently, another three-dimensional CY geometry was identified at 5PM 1SF order in the dissipative sector~\cite{Klemm:2024wtd}. Notably, the full 5PM 1SF dissipative sector was recently computed in ref.~\cite{Driesse:2024feo}, confirming that the corresponding CY functions appear in the physical observables. However, a complete calculation at 5PM 2SF order is still missing and is considered significantly more challenging.

In this paper, we complete the classification of all geometries and special functions that appear in the 5PM correction to all SF orders, using the method developed in refs.~\cite{Frellesvig:2023bbf,Frellesvig:2024zph}. In particular, the approach is based on calculating the leading singularity~\cite{Cachazo:2008vp,Arkani-Hamed:2010pyv} of all classically contributing Feynman integrals, a task that can be easily performed using the loop-by-loop Baikov representation~\cite{Frellesvig:2017aai,Frellesvig:2024ymq}. Due to the linearized propagators characteristic of the PM expansion, these leading singularities satisfy numerous linear relations, which significantly reduce the number of genuinely independent integrals that need to be analyzed. Additionally, we identify certain subgraphs whose presence implies that the corresponding integral topology is fully reducible to subsectors.
As a consequence, the number of independent integral topologies at 5PM order is reduced from 16,596 to only 70. We then compute the leading singularities for one integral in these remaining integral topologies, thereby revealing the underlying geometries; see tables~\ref{tab: 4-loop results 0SF and 1SF} and~\ref{tab: 4-loop results 2SF} for a summary of our results. In total, we identify 8 independent integral topologies that depend on non-trivial geometries, which are listed in fig.~\ref{tab:reduction_diagrams}. Those geometries characterize the class of special functions that will be necessary to perform a full calculation of gravitational-wave observables at 5PM order.

The remainder of this paper is structured as follows. In sec.~\ref{sec:review}, we review the method that we use for our analysis of Feynman integral geometries. Concretely, in sec.~\ref{sec:review_PM_integrals}, we provide a brief review of the PM expansion of classical gravity and the integrals that may be found therein. In sec.~\ref{sec:review_DE_LS}, we present the methods of differential equations and leading singularities, and discuss their relation with the underlying geometries. Then, in sec.~\ref{sec:review_relating_geometries}, we gather the linear relations obeyed by the leading singularities of PM integrals, which we use to reduce the size of our analysis. We list our results in sec.~\ref{sec:results}, with the primary focus being the integral topologies depending on non-trivial geometries. Lastly, in sec.~\ref{sec:conclusions}, we conclude and discuss further research directions. Attached to the arXiv submission of this paper, we provide \texttt{Mathematica} notebooks containing the Baikov representation and the full calculation of the 70 independent leading singularities for both conservative and dissipative contributions, as well as the corresponding results.

\begin{figure}[t]
    \centering
    \begin{tabular}{c @{\hspace{1cm}} c @{\hspace{1cm}} c @{\hspace{1cm}} c @{\hspace{1cm}} c}
        1SF: & \begin{tikzpicture}[baseline={([yshift=-0.1cm]current bounding box.center)}] 
            \node[] (a) at (-0.5,0) {};
            \node[] (a1) at (0,0) {};
            \node[] (a2) at (1,0) {};
            \node[] (a3) at (1.5,0) {};
            \node[] (b) at (0.5,-0.5) {};
            \node[] (c) at (0,-1) {};
            \node[] (c1) at (1,-1) {};
            \draw[line width=0.15mm] (c.center) --  (a.center);
            \draw[line width=0.15mm] (a1.center) --  (b.center);
            \draw[line width=0.15mm] (c.center) --  (b.center);
            \draw[line width=0.15mm] (b.center) --  (c1.center);
            \draw[line width=0.5mm] (a.center) --  (a3.center);
            \draw[line width=0.15mm] (b.center) --  (a2.center);
            \draw[line width=0.5mm] (c.center) --  (c1.center);
            \draw[line width=0.5mm] (-0.65,0) -- (a.center);
            \draw[line width=0.5mm] (a3.center) -- (1.65,0);
            \draw[line width=0.5mm] (-0.65,-1) -- (c.center);
            \draw[line width=0.15mm] (c1.center) --  (a3.center);
            \draw[line width=0.5mm] (c1.center) -- (1.65,-1);
        \end{tikzpicture} & \begin{tikzpicture}[baseline={([yshift=-0.1cm]current bounding box.center)}] 
	\node[] (a) at (0,0) {};
	\node[] (a1) at (0.5,0) {};
	\node[] (a2) at (1,0) {};
    \node[] (a3) at (1.5,0) {};
	\node[] (b) at (0,-0.5) {};
	\node[] (b1) at (0.75,-0.5) {};
	\node[] (b2) at (0.75,-0.5) {};
    \node[] (b3) at (1.5,-0.5) {};
	\node[] (c) at (0,-1) {};
	\node[] (c1) at (0.5,-1) {};
	\node[] (c2) at (1,-1) {};
    \node[] (c3) at (1.5,-1) {};
	\node[] (p1) at ($(a)+(-0.2,0)$) {};
	\node[] (p2) at ($(c)+(-0.2,0)$) {};
	\node[] (p3) at ($(c3)+(0.2,0)$) {};
	\node[] (p4) at ($(a3)+(0.2,0)$) {};
    \draw[line width=0.5mm] (p1.center) -- (p4.center);
	\draw[line width=0.5mm] (p2.center) -- (p3.center);
    \draw[line width=0.15mm] (c.center) -- (a.center);
	\draw[line width=0.15mm] (b.center) -- (b1.center);
	\draw[line width=0.15mm] (b1.center) -- (b2.center);
    \draw[line width=0.15mm] (b1.center) -- (a1.center);
    \draw[line width=0.15mm] (b2.center) -- (c3.center);
	\draw[line width=0.15mm] (b2.center) -- (a2.center);
    \draw[line width=0.15mm] (a3.center) -- (c3.center);
    \draw[line width=0.5mm] (a2.center) --(a3.center);
\end{tikzpicture} & \begin{tikzpicture}[baseline={([yshift=-0.1cm]current bounding box.center)}] 
	\node[] (a) at (0,0) {};
	\node[] (a1) at (0.5,0) {};
	\node[] (a2) at (1,0) {};
    \node[] (a3) at (1.5,0) {};
	\node[] (b) at (0,-0.5) {};
	\node[] (b1) at (0.5,-0.5) {};
	\node[] (b2) at (1,-0.5) {};
    \node[] (b3) at (1.5,-0.5) {};
	\node[] (c) at (0,-1) {};
	\node[] (c1) at (0.5,-1) {};
	\node[] (c2) at (1,-1) {};
    \node[] (c3) at (1.5,-1) {};
	\node[] (p1) at ($(a)+(-0.2,0)$) {};
	\node[] (p2) at ($(c)+(-0.2,0)$) {};
	\node[] (p3) at ($(c3)+(0.2,0)$) {};
	\node[] (p4) at ($(a3)+(0.2,0)$) {};
    \draw[line width=0.5mm] (p1.center) -- (p4.center);
	\draw[line width=0.5mm] (p2.center) -- (p3.center);
    \draw[line width=0.15mm] (c.center) -- (a.center);
	\draw[line width=0.15mm] (b.center) -- (b1.center);
	\draw[line width=0.15mm] (b1.center) -- (b2.center);
    \draw[line width=0.15mm] (b.center) -- (a1.center);
    \draw[line width=0.15mm] (b2.center) -- (c3.center);
	\draw[line width=0.15mm] (b2.center) -- (a2.center);
    \draw[line width=0.15mm] (a3.center) -- (c3.center);
     \draw[line width=0.5mm] (a2.center) -- (a3.center);
\end{tikzpicture} & \begin{tikzpicture}[baseline={([yshift=-0.1cm]current bounding box.center)}] 
	\node[] (a) at (0,0) {};
	\node[] (a1) at (0.5,0) {};
	\node[] (a2) at (1,0) {};
    \node[] (a3) at (1.5,0) {};
	\node[] (b) at (0,-0.5) {};
	\node[] (b1) at (0.5,-0.5) {};
	\node[] (b2) at (1,-0.5) {};
    \node[] (b3) at (1.5,-0.5) {};
	\node[] (c) at (0,-1) {};
	\node[] (c1) at (0.5,-1) {};
	\node[] (c2) at (1,-1) {};
    \node[] (c3) at (1.5,-1) {};
	\node[] (p1) at ($(a)+(-0.2,0)$) {};
	\node[] (p2) at ($(c)+(-0.2,0)$) {};
	\node[] (p3) at ($(c3)+(0.2,0)$) {};
	\node[] (p4) at ($(a3)+(0.2,0)$) {};
    \draw[line width=0.5mm] (p1.center) -- (p4.center);
	\draw[line width=0.5mm] (p2.center) -- (p3.center);
    \draw[line width=0.15mm] (c.center) -- (a.center);
	\draw[line width=0.15mm] (b.center) -- (b1.center);
	\draw[line width=0.15mm] (b1.center) -- (b2.center);
    \draw[line width=0.15mm] (b.center) -- (a1.center);
    \draw[line width=0.15mm] (b2.center) -- (b3.center);
	\draw[line width=0.15mm] (b3.center) -- (a2.center);
    \draw[line width=0.15mm] (a3.center) -- (c3.center);
\end{tikzpicture}
        \\
        & \small $\text{CY}_3$ & \small K3 & \small K3 & \small K3 \\[0.5cm]
        
        2SF: & \begin{tikzpicture}[baseline={([yshift=-0.1cm]current bounding box.center)}] 
            \node[] (a) at (0,0) {};
            \node[] (a1) at (0.75,0) {};
            \node[] (a2) at (1.5,0) {};
            \node[] (b) at (0,-0.5) {};
            \node[] (b1) at (0.5,-0.5) {};
            \node[] (b2) at (1.5,-0.5) {};
            \node[] (c) at (0,-1) {};
            \node[] (c1) at (0.75,-1) {};
            \node[] (c2) at (1.5,-1) {};
            \node[] (p1) at ($(a)+(-0.2,0)$) {};
            \node[] (p2) at ($(c)+(-0.2,0)$) {};
            \node[] (p3) at ($(c2)+(0.2,0)$) {};
            \node[] (p4) at ($(a2)+(0.2,0)$) {};
            \draw[line width=0.15mm] (c.center) -- (a.center);
            \draw[line width=0.15mm] (b.center) -- (b2.center);
            \draw[line width=0.15mm] (c2.center) -- (a2.center);
            \draw[line width=0.15mm] (b2.center) -- (c1.center);
            \draw[line width=0.15mm] (b.center) -- (a1.center);
            \draw[line width=0.5mm] (p1.center) -- (p4.center);
            \draw[line width=0.5mm] (p2.center) -- (p3.center);
        \end{tikzpicture} & \begin{tikzpicture}[baseline={([yshift=-0.1cm]current bounding box.center)}] 
	\node[] (a) at (0,0) {};
	\node[] (a1) at (0.5,0) {};
	\node[] (a2) at (1,0) {};
    \node[] (a3) at (1.5,0) {};
	\node[] (b) at (0,-0.5) {};
	\node[] (b1) at (0.5,-0.5) {};
	\node[] (b2) at (1,-0.5) {};
    \node[] (b3) at (1.5,-0.5) {};
	\node[] (c) at (0,-1) {};
	\node[] (c1) at (0.5,-1) {};
	\node[] (c2) at (0.75,-1) {};
    \node[] (c3) at (1.5,-1) {};
	\node[] (p1) at ($(a)+(-0.2,0)$) {};
	\node[] (p2) at ($(c)+(-0.2,0)$) {};
	\node[] (p3) at ($(c3)+(0.2,0)$) {};
	\node[] (p4) at ($(a3)+(0.2,0)$) {};
    \draw[line width=0.5mm] (p1.center) -- (p4.center);
	\draw[line width=0.5mm] (p2.center) -- (p3.center);
    \draw[line width=0.15mm] (c.center) -- (a.center);
	\draw[line width=0.15mm] (b.center) -- (a2.center);
    \draw[line width=0.15mm] (b.center) -- (c3.center);
    \draw[line width=0.15mm] (c2.center) -- (b.center);
    \draw[line width=0.15mm] (a3.center) -- (c3.center);
\end{tikzpicture} & \begin{tikzpicture}[baseline={([yshift=-0.235cm]current bounding box.center)}]
    \node[] (a) at (0,0) {};
    \node[] (a1) at (0.8,0) {};
    \node[] (a2) at (1,0) {};
    \node[] (a3) at (1.5,0) {};
    \node[] (b1) at (1,-0.75) {};
    \node[] (c) at (0,-1) {};
    \node[] (c1) at (0.8,-1) {};
    \node[] (c2) at (1.25,-1) {};
    \node[] (c3) at (1.5,-1) {};
    \node[] (p1) at ($(a)+(-0.2,0)$) {};
	\node[] (p2) at ($(c)+(-0.2,0)$) {};
	\node[] (p3) at ($(c3)+(0.2,0)$) {};
	\node[] (p4) at ($(a3)+(0.2,0)$) {};
    \draw[line width=0.5mm] (p1.center) -- (p4.center);
	\draw[line width=0.5mm] (p2.center) -- (p3.center);
    \draw[line width=0.15mm] (a.center) -- (c.center);
    \draw[line width=0.15mm] (a3.center) -- (c1.center);
    \draw[line width=0.15mm] (a1.center) -- (c3.center);
    \draw[line width=0.15mm] (0.8,0) arc (0:180:0.4);
\end{tikzpicture} & \begin{tikzpicture}[baseline={([yshift=-0.1cm]current bounding box.center)}] 
	\node[] (a) at (0,0) {};
	\node[] (a1) at (0.75,0) {};
	\node[] (a2) at (1.5,0) {};
	\node[] (c) at (0,-1) {};
	\node[] (c1) at (0.75,-1) {};
	\node[] (c2) at (1.5,-1) {};
	\node[] (p1) at ($(a)+(-0.2,0)$) {};
	\node[] (p2) at ($(c)+(-0.2,0)$) {};
	\node[] (p3) at ($(c2)+(0.2,0)$) {};
	\node[] (p4) at ($(a2)+(0.2,0)$) {};
	\draw[line width=0.15mm] (a.center) -- (c.center);
	\draw[line width=0.15mm] (a1.center) -- (c1.center);
	\draw[line width=0.15mm] (a2.center) -- (c2.center);
	\draw[line width=0.15mm] (a.center) -- (c2.center);
	\draw[line width=0.5mm] (p1.center) -- (p4.center);
	\draw[line width=0.5mm] (p2.center) -- (p3.center);
\end{tikzpicture}
        \\
        & \small $\text{CY}'_3$ & \small K3 & \small K3 & \small $\text{K3}'$
    \end{tabular}
    \caption{Independent four-loop PM Feynman integral topologies which contain non-trivial geometries. In the first row (1SF order), the first topology depends on a three-dimensional Calabi--Yau geometry, while the remaining ones depend on the same K3 surface that appeared at three loops~\cite{Ruf:2021egk,Dlapa:2022wdu,Frellesvig:2024zph}. In the second row (2SF order), the first topology depends on a different three-dimensional Calabi--Yau geometry, the next two involve the same K3 surface as in 1SF order, and the last one involves a different K3 surface.}
    \label{fig: non-trivial_diagrams_4loop}
\end{figure}

\section{Post-Minkowskian integrals and leading singularities}
\label{sec:review}

In this section, we briefly review the post-Minkowskian expansion for the two-body problem in classical gravity and present the details of our method for analyzing Feynman integral geometries. First, in sec.~\ref{sec:review_PM_integrals}, we introduce the kinematics of the process and discuss the type of Feynman integrals that contribute to the classical scattering of non-spinning black holes; see refs.~\cite{Bjerrum-Bohr:2022blt,Buonanno:2022pgc,Frellesvig:2024zph} for more thorough introductions. Then, in sec.~\ref{sec:review_DE_LS}, we review the methods of differential equations~\cite{Kotikov:1990kg} and leading singularities~\cite{Cachazo:2008vp,Arkani-Hamed:2010pyv}, and highlight how they can be used to detect and characterize the underlying Feynman integral geometries. In particular, we review the loop-by-loop Baikov representation~\cite{Frellesvig:2017aai,Frellesvig:2024ymq}, which can be used to efficiently compute the leading singularity of PM integrals. Subsequently, in sec.~\ref{sec:review_relating_geometries}, we present the relations that the leading singularities of PM Feynman integrals satisfy, and which allow us to connect their geometries across different loop orders and integral topologies, considerably reducing the size of our analysis.

\subsection{Review of post-Minkowskian Feynman integrals}
\label{sec:review_PM_integrals} 

To describe the scattering of two compact astronomical objects, such as black holes and neutron stars, we can assume a long-distance regime and regard the impact parameter $|b|$ to be much bigger than the Schwarzschild radii of the black holes, i.e.~$r_s \ll |b|$. Under this hierarchy of scales, we can model the black holes in terms of massive scalars and adopt an effective field theory approach~\cite{Cheung:2018wkq}. The internal structure and the spin degrees of freedom can be taken into consideration via a systematic expansion that only effects the numerators of the Feynman integrals, and thus does not influence the occurring special functions and geometries. 

\begin{figure}
\centering
\begin{tikzpicture}[baseline={([yshift=-0.1cm]current bounding box.center)}, scale=1.2] 
	\node[] (a) at (0,0) {};
	\node[] (b) at (0,-1) {};
	\node[label=left:{$p_1$}] (p1) at ($(a)+(-1,0)$) {};
	\node[label=left:{$p_2$}] (p2) at ($(b)+(-1,0)$) {};
	\node[label=right:{$p_3={-}p_2{-}q$}] (p3) at ($(b)+(1,0)$) {};
	\node[label=right:{$p_4={-}p_1{+}q$}] (p4) at ($(a)+(1,0)$) {};
	\draw[line width=0.15mm, postaction={decorate}] (b.center) -- node[sloped, allow upside down, label={[xshift=0.75cm, yshift=0cm]$q$}] {\midarrow} (a.center);
	\draw[line width=0.5mm, postaction={decorate}] (a.center) -- node[sloped, allow upside down] {\midarrow} (p1.center);
	\draw[line width=0.5mm, postaction={decorate}] (a.center) -- node[sloped, allow upside down] {\midarrow} (p4.center);
	\draw[line width=0.5mm, postaction={decorate}] (b.center) -- node[sloped, allow upside down] {\midarrow} (p2.center);
	\draw[line width=0.5mm, postaction={decorate}] (b.center) -- node[sloped, allow upside down] {\midarrow} (p3.center);
\end{tikzpicture}
\caption{Kinematics for the scattering process, exemplified by the tree-level exchange. The massive scalars are depicted by thick lines, while the graviton propagator is denoted by a thin line.}
\label{fig:set-up}
\end{figure}
The kinematics of the scattering process are provided in fig.~\ref{fig:set-up}, which represents the tree-level diagram (1PM order). In particular, the black holes are modeled by two massive scalars, which have masses and momenta given by $m_i$ and $p_i$, respectively. Under the classical limit, the momentum transfer $|q| \sim 1/|b|$ transferred between the two scalars becomes small, i.e.~$|q| \ll 1$, which leads to the so-called soft expansion~\cite{Neill:2013wsa}. In this limit, the matter propagators can be expanded and become linearized,
\begin{equation}
\raisebox{0.2cm}{
\begin{tikzpicture}[baseline={([yshift=-0.1cm]current bounding box.center)}] 
	\node[] (a) at (0,0) {};
	\node[] (p4) at ($(a)+(2,0)$) {};
	\draw[line width=0.5mm, postaction={decorate}] (a.center) -- node[sloped, allow upside down, label={[xshift=0cm, yshift=-0.2cm]$k$}] {\midarrow} (p4.center);
\end{tikzpicture}} = \frac{1}{m_i}\frac{1}{2u_i\cdot k}+ \mathcal{O}(q^2)\,,
\end{equation}
where $u_i^\mu=p_i^\mu/m_i+\mathcal{O}(|q|)$ is the soft four-velocity, satisfying $u_i^2=1$ and $u_i \cdot q=0$. Since the mass dependence factors out, the classical scattering amplitude thus depends only on the scale $q^2$, which can be fixed to $q^2=-1$ and recovered with dimensional analysis, as well as on a single dimensionless variable
\begin{equation}
y=u_1\cdot u_2=\frac{p_1\cdot p_2}{m_1 m_2}+\mathcal{O}(q^2)\,.
\end{equation}
This parameter is typically rewritten as $y=\frac{1+x^2}{2x}$ to rationalize the square root $\sqrt{y^2-1}$ that occurs throughout the calculations.

While the perturbative expansion for this $2 \to 2$ scattering process yields a plethora of Feynman integrals, only a small subset actually contributes in the classical limit. At $(L+1)$ PM order, which corresponds to $L$ loops, integrals that scale as $|q|^a$ with $a>L-2$ correspond to quantum corrections and can be discarded, whereas integrals with a power scale $a=L-2$ provide the classical contribution~\cite{Neill:2013wsa}. In addition, there are also superclassical integrals, with $a<L-2$. These contributions cancel against products of lower-loop integrals~\cite{Bern:2019crd}, and thus cannot introduce new functions. We will nevertheless take them into account in our analysis for completeness. In the soft expansion, linearized scalar propagators scale as $1/|q|$, graviton propagators scale as $1/|q|^2$, vertices involving scalars scale as $1$ and pure graviton vertices scale as $|q|^2$. Moreover, each loop integral contributes with $|q|^4$. With these rules, one can easily verify which integrals contribute in the classical limit. In particular, integrals with closed graviton loops are always quantum, such that classical integrals need to contain one scalar propagator per loop. Finally, integrals with a direct contact between the scalar lines are dimensionless and thus vanish in dimensional regularization. 

Each correction in the PM expansion can moreover be naturally organized according to the self-force (SF) expansion~\cite{Mino:1996nk,Quinn:1996am,Poisson:2011nh,Barack:2018yvs}, such that an integral with $n_i$ scalar propagators on the line $u_i$ contributes to $\text{min}(n_1,n_2)$ SF order. Explicitly, at $L$ loops, classical PM Feynman integrals with $0$ scalar propagators on a line contribute to 0SF order, those with $1$ scalar propagator on a line contribute to 1SF order, up to $\lfloor \frac{L}{2} \rfloor$ SF order. Lastly, since the scalar propagators become linearized under the soft expansion, PM Feynman integrals have a definite parity under the transformation $u_i\to -u_i$ for both $i=1,2$. Concretely, the integrals can either be even or odd under parity.\footnote{Note that our definition of even and odd differs from the one typically used in the worldline formalism, where $L$ linearized propagators at $L$ loops are replaced by delta functions.} While one of the two parities suffices to calculate conservative corrections to gravitational-wave observables such as the impulse~\cite{Kalin:2020fhe,Dlapa:2023hsl,Mogull:2020sak}, we are interested in both conservative and dissipative contributions and thus need to consider both parities for each integral topology; see also the discussion in ref.~\cite{Frellesvig:2024zph}.

\subsection{Identifying non-trivial geometries}
\label{sec:review_DE_LS}

In this subsection, we briefly review the two main methods that can be used to detect and identify the geometry associated with a given Feynman integral, namely differential equations~\cite{Kotikov:1990kg} and leading singularities~\cite{Cachazo:2008vp,Arkani-Hamed:2010pyv}.

The first method to detect geometries is based on the differential equations that Feynman integrals satisfy~\cite{Kotikov:1990kg}. In particular, after solving the integration-by-parts identities (IBPs)~\cite{Chetyrkin:1981qh}, all integrals within a topology can be reduced to a minimal set of Feynman integrals, the so-called master integrals. Organizing them in a vector $\vec{\mathcal{I}}$ and taking a derivative with respect to a kinematic variable, e.g.\ $x$, we obtain a system of coupled first-order differential equations $\partial_x \vec{\mathcal{I}}=A \, \vec{\mathcal{I}}$. Here, $A$ is a matrix with rational entries which depend on the kinematics and the space-time dimension $D=4-2\varepsilon$. In particular, the parity-splitting of PM integrals generated by the linearized matter propagators separates the differential equation into two decoupled blocks, see e.g.\ refs.~\cite{Parra-Martinez:2020dzs,DiVecchia:2021bdo,Herrmann:2021tct}, which we assume do not decouple any further~\cite{Frellesvig:2024zph}.

By changing the basis of master integrals to a derivative basis, we can reduce the system to a single higher-order differential equation for one of the master integrals in each parity sector. This equation takes the form
\begin{equation}
\label{eq: Picard-Fuchs_eq}
\mathcal{L}_n \mathcal{I}_i = \left( \frac{d^n}{d x^n} + \sum_{j=0}^{n-1} c_j(x,\varepsilon) \frac{d^j}{d x^j} \right) \mathcal{I}_i = \text{inhomogeneity},
\end{equation}
where the operator $\mathcal{L}_n$ is known as the Picard--Fuchs operator of the master integral $\mathcal{I}_i$, with the $c_j(x,\varepsilon)$ being rational functions. The inhomogeneity stems from master integrals of subsectors, which correspond to integrals where some of the propagators are absent, i.e.\ the corresponding edges in the graph are pinched.

Typically, the Picard--Fuchs operator restricted to $D=4$ dimensions has a rational factorization into a product of (possibly distinct) lower-order operators, $\mathcal{L}_n = \prod_j \mathcal{L}_{n_j}^{(j)}$, where the $n_j$ sum to $n$. Importantly, each operator $\mathcal{L}_{n_j}^{(j)}$ in the factorization corresponds to a geometry characteristic of the Feynman integral, and determines the space of functions that the integral $\mathcal{I}_i$ evaluates to~\cite{Adams:2017tga}. Concretely, if the Picard--Fuchs operator completely factorizes into a product $\mathcal{L}^{(1)}_1 \cdots \mathcal{L}^{(n)}_1$ of operators of order one, and the same is true iteratively for all subsectors in the inhomogeneity, then the integral $\mathcal{I}_i$ admits a differential equation in $d \log$ form, with a solution often being expressible in terms of multiple polylogarithms. By contrast, if the factorization contains an irreducible operator $\mathcal{L}^{(j)}_{n_j}$ of order greater than one, it indicates a dependence on a non-trivial geometry, such that the solution for the master integral lies beyond the realm of polylogarithms. In particular, provided that the Picard--Fuchs operator fulfills a number of additional criteria~\cite{Bogner:2013kvr}, it typically corresponds to an integral over an $(n_j-1)$-dimensional Calabi--Yau geometry; see e.g.\ ref.~\cite{Frellesvig:2024rea} for a discussion in the context of PM integrals. 

While analyzing the Picard--Fuchs operator of a master integral provides definitive evidence for the presence of non-trivial geometries, it comes with a downside. Concretely, in order to obtain the operator, one needs to first perform the IBP reduction into master integrals for each integral sector. Although there exists an algorithmic solution~\cite{Laporta:2000dsw}, this step represents a genuine bottleneck when applied to state-of-the-art integrals with a level of complexity comparable to the ones discussed in this paper, even with the availability of highly optimized computer implementations~\cite{Smirnov:2023yhb,vonManteuffel:2012np,Lee:2013mka,Klappert:2020nbg}. Consequently, we will mostly use a different approach to detect non-trivial geometries; see, however, refs.~\cite{JohannQCDmeetsGravity,Driesse:2024xad,Guan:2024byi,Bern:2024adl,vonHippel:2025okr,Song:2025pwy,Zeng:2025xbh,Lange:2025fba} for recent improvements of IBP reduction algorithms. Similarly, we refer to refs.~\cite{Brunello:2023fef,Frellesvig:2024swj} for a method to obtain PM integral relations based on intersection theory~\cite{Mastrolia:2018uzb,Frellesvig:2019uqt}, as well as refs.~\cite{Lairez:2022zkj,delaCruz:2024xit} for an alternative derivation of Picard--Fuchs operators via Griffiths--Dwork reduction, which does not rely on IBPs.

As an alternative to differential equations, we can analyze the leading singularity (LS) of the integral~\cite{Cachazo:2008vp,Arkani-Hamed:2010pyv}. The leading singularity corresponds to taking the maximal cut, where all propagators go on-shell, $\frac{i}{Q_i^2 - m_i^2} \to \delta(Q_i^2 - m_i^2)$, as well as taking any possible further residues that can be done at the level of rational functions. Insisting on only taking residues at the level of rational functions deviates from the definition used in other parts of the literature but is crucial for our approach; see also the discussion in the next two paragraphs and in ref.\ \cite{Frellesvig:2024zph}. In this work, we use the notion of generalized cuts~\cite{Britto:2024mna}, such that we compute the residue at the point $Q_i^2=m_i^2$ by deforming the integration contour to encircle the pole. Since the action of cutting a propagator tends to commute with taking derivatives, the leading singularity provides a solution to the differential equation that Feynman integrals satisfy. Concretely, since the maximal cut annihilates all subsectors, the leading singularity corresponds to a solution of the Picard--Fuchs operator, i.e.~of the homogeneous part of eq.~\eqref{eq: Picard-Fuchs_eq}. Therefore, it may also be used to study Feynman integral geometries, see e.g.~refs.~\cite{Primo:2016ebd,Bosma:2017ens}, allowing us to analyze one integral sector at a time.

The outcome of computing the leading singularity is either an algebraic function of the kinematics or a transcendental integral, which indicates a non-trivial geometry. In particular, if the leading singularities of an integral and all of its subsectors are algebraic, then the integral has a $d \log$ form. Otherwise, the integral depends on a non-trivial geometry, and the result of its leading singularity can often be brought to the form
\begin{align}
\label{eq: LS_definition}
\LS \left( \mathcal{I} \right) &\propto \int \! \frac{d^n \vec{z}}{\sqrt{P_m(\vec{z})}}\,,
\end{align}
where $P_m(\vec{z})$ is a polynomial of total degree $m$ in the $n$ integration variables $z_i$. If $\sqrt{P_m(\vec{z})}$ cannot be rationalized by a suitable change of variables or avoided by different steps earlier in the calculation, and if moreover $m = 2(n+1)$, then eq.~\eqref{eq: LS_definition} shows that the integral topology of $\mathcal{I}$ depends on an $n$-dimensional Calabi--Yau geometry; see refs.~\cite{Hubsch:1992nu,Bourjaily:2019hmc} for further details. While it is hard to prove that no rationalization or alternative sequence of steps exists for the leading singularity, the result of the leading singularity being an algebraic function completely rules out the possibility of such non-trivial geometries entering through a given sector.
 Therefore, since computing the leading singularity is much more efficient than solving the IBP relations to obtain the differential equations, we will only compute the Picard--Fuchs operators for cases involving non-trivial geometries as a cross-check.
 
Crucially, the procedure of taking residues in further poles after performing the maximal cut is only indicative of the geometry if all square roots involving that integration variable have been rationalized, i.e.\ when taking only further residues that can be done at the level of rational functions.
 The reasoning behind this restriction is discussed in detail in ref.\ \cite{Frellesvig:2024zph}, but its necessity is illustrated by elliptic integrals of the third kind. These integrals schematically take the form 
\begin{equation}
\label{eq: third kind}
\int \frac{dx}{(x-a) \sqrt{P_4(x)}}\,,
\end{equation}
and naively taking the residue at $x=a$ could lead to the mistaken conclusion that no elliptic curve is present.

In order to compute the leading singularity for PM Feynman integrals, we will use the Baikov representation~\cite{Baikov:1996iu}, where the propagators correspond directly to the integration variables, thus trivializing the maximal cut. We will not introduce the by-now well-known Baikov representation in further detail, though; see refs.~\cite{Weinzierl:2022eaz,Frellesvig:2024ymq} for an introduction as well as ref.~\cite{Frellesvig:2024zph} for a recent application in the PM expansion up to three loops. In particular, we will use the loop-by-loop (LBL) Baikov representation~\cite{Frellesvig:2017aai}, implemented in the \texttt{\textup{BaikovPackage}} package~\cite{Frellesvig:2024ymq} in \texttt{\textup{Mathematica}}. In this case, the standard Baikov representation is used one loop at a time, which reduces the number of variables in the integral representation, see refs.~\cite{Frellesvig:2024zph,Frellesvig:2024ymq} for further details. For us, the relevant part is the result of taking the maximal cut in the LBL Baikov representation, which for an $L$-loop integral takes the form
\begin{equation}
\mathcal{I}_{\text{max-cut}} \propto \int_{\mathcal{C}} d^{\nISP} \vec{z} \ \mathcal{N}(\vec{z}) \ {\mathcal{B}_1(\vec{z})}^{\rho_1} \cdots \, {\mathcal{B}_{2L-1}(\vec{z})}^{\rho_{2L-1}}\,.
\label{eq: cut_Baikov}
\end{equation}
In this expression, $\rho_i$ depend on the space-time dimension $D$ and the number of independent external momenta with respect to each loop. For $D=4$, the $\rho_i$ can be either integer or half-integer, such that multiple square roots in the integration variables can appear simultaneously in the result of the maximal cut. In addition, the $\mathcal{B}_ i(\vec{z})$ are polynomials, which correspond to the so-called Baikov polynomials after the maximal cut. The integration is performed over a region (or chamber) $\mathcal{C}$ bounded by the vanishing condition for these Baikov polynomials, such that the powers $\rho$ ensure that the integration is regulated. 

As can be seen, eq.~\eqref{eq: cut_Baikov} involves $\nISP$ integrals, where $\nISP$ corresponds to the number of irreducible scalar products (ISPs) that can be formed between loop momenta and external momenta and which cannot be found in the propagators of the integral. Specifically, in the LBL Baikov parametrization we have
\begin{equation}
\label{eq: n_ISP}
\nISP = L + \sum_{i=1}^L E_i - \nint\,,
\end{equation}
where $\nint$ denotes the number of propagators, and $E_i$ are the number of independent external momenta with respect to each loop. Lastly, in eq.~\eqref{eq: cut_Baikov} we also include $\mathcal{N}(\vec{z})$, which denotes a generic polynomial of the ISPs. Such a numerator factor may arise from the Feynman rules or, more relevant in this context, from cuts of propagators raised to higher powers. In such a case, the maximal cut results in derivatives of the Baikov polynomials, which may be absorbed in $\mathcal{N}(\vec{z})$.

As can be clearly seen from eq.~\eqref{eq: cut_Baikov}, if the leading singularity involves a non-trivial geometry, it must arise as a transcendental integral over the ISPs, such as eq.~\eqref{eq: LS_definition}. However, it often occurs that the dimension of the actual Feynman integral geometry is much smaller than $\nISP$. In general, this is partly due to the fact that, after taking the maximal cut, further poles are exposed. The leading singularity hence corresponds to taking the residue also at those poles, which may reduce the complexity of the transcendental integral. In practice, this can be efficiently done using the {\texttt{\textup{LeadingSingularities}}} command from the \texttt{\textup{DlogBasis}} package~\cite{Henn:2020lye} in \texttt{\textup{Mathematica}}. Importantly, such poles are frequently only apparent after performing suitable changes of variables in the $\vec{z}$ variables. Likewise, obtaining a result with a single square root of the form of eq.~\eqref{eq: LS_definition} usually involves rationalizing multiple square roots in eq.~\eqref{eq: cut_Baikov} via changes of variables.

In this work, we found it sufficient to invoke two different variable transformations to analyze the leading singularities of PM integrals at four loops. In particular, these transformations coincide with the changes of variables used in ref.~\cite{Frellesvig:2024zph} for the analysis up to three loops. First, for square roots of quadratic polynomials $\sqrt{(z_i-r_1)(z_i-r_2)}\,$, where $r_j$ are the roots, we can use the following change of variables to $t_i$~\cite{Bonciani:2010ms,Adams:2018yfj}:
\vspace*{-0.2cm}
\begin{equation}
\label{eq: change_of_variables_(z-r1)(z-r2)}
z_i = r_1-\frac{(r_2-r_1)(1-t_i)^2}{4t_i}\,.
\end{equation}
Note that for $r_1=1$, $r_2=-1$, this transformation corresponds to the change of variables $y=(1+x^2)/2x$ used in sec.~\ref{sec:review_PM_integrals} to rationalize the square root $\sqrt{y^2-1}$ in the kinematics.
The same change of variables works for rationalizing two square roots of linear polynomials simultaneously, i.e.\ $\sqrt{(z_i-r_1)}\sqrt{(z_i-r_2)}$.
Similarly, for $\sqrt{z_i-r^2}$ we can use the change of variables
\begin{equation}
\label{eq: change_of_variables_(z-c_squared)}
z_i = \frac{1-2i t_i r}{t_i^2}\,,
\end{equation}
which we see depends only linearly on $r$.

There are nevertheless cases in which we need to perform the integration between two zeros of the Baikov polynomials, since there are no changes of variables that expose further poles in which to take residues; see ref.~\cite{Frellesvig:2024zph} for details. We will encounter one such example in sec.~\ref{sec:CY2}. 

Lastly, throughout this work, we assume that all integrals within an integral topology couple to each other in the differential equation for a given parity sector. Under this assumption, analyzing the leading singularity of a single integral per parity sector and integral topology is sufficient to characterize the underlying geometry; see also the discussion in ref.~\cite{Frellesvig:2024zph} and around eq.~\eqref{eq: third kind}.

\subsection{Relating geometries across integral topologies}
\label{sec:review_relating_geometries}

While it would be possible to study the geometry of each individual Feynman integral topology that contributes to the classical limit, in ref.~\cite{Frellesvig:2024zph} it was found that only an independent subset needs to be analyzed in detail. In particular, there exist relations among leading singularities which allow us to connect Feynman integral geometries across different integral topologies and loop orders. Moreover, large classes of integral topologies are found to contain no master integrals since they can be fully reduced to subsectors via IBP relations. Both types of relations drastically reduce the number of independent integral topologies that need to be considered, see table~\ref{tab:reduction_diagrams}, and therefore significantly accelerate our analysis. While we refer the reader to ref.~\cite{Frellesvig:2024zph} for more details, let us briefly summarize these simplifying relations.

First, for superclassical integral topologies containing isolated graviton lines that attach to the matter lines via cubic vertices, we have the relation
\begin{equation}
\label{eq: reduction_superclassical}
\LS \left(
\begin{tikzpicture}[baseline={([yshift=-0.1cm]current bounding box.center)}] 
	\node[] (a) at (0,0) {};
	\node[] (a1) at (0.4,0) {};
	\node[] (a2) at (1.5,0) {};
	\node[] (a3) at (2.6,0) {};
	\node[] (a4) at (3,0) {};
	\node[] (b) at (0,-1.2) {};
	\node[] (b1) at (0.4,-1.2) {};
	\node[] (b2) at (1.5,-1.2) {};
	\node[] (b3) at (2.6,-1.2) {};
	\node[] (b4) at (3,-1.2) {};
	\node[] (p1) at ($(a)+(-0.3,0)$) {};
	\node[] (p2) at ($(b)+(-0.3,0)$) {};
	\node[] (p3) at ($(b4)+(0.3,0)$) {};
	\node[] (p4) at ($(a4)+(0.3,0)$) {};
	\draw[line width=0.15mm] (a2.center) -- (b2.center);
	\draw[line width=0.5mm] (p1.center) -- (0.4,0);
	\draw[line width=0.5mm] (0.4,0) -- (2.6,0);
	\draw[line width=0.5mm] (2.6,0) -- (p4.center);
	\draw[line width=0.5mm] (p2.center) -- (0.4,-1.2);
	\draw[line width=0.5mm] (0.4,-1.2) -- (2.6,-1.2);
	\draw[line width=0.5mm] (2.6,-1.2) -- (p3.center);
	\fill[gray!50] ($(a1)+(0,-0.6)$) ellipse (0.4 and 0.7);
	\draw ($(a1)+(0,-0.6)$) ellipse (0.4 and 0.7);
	\fill[gray!50] ($(a3)+(0,-0.6)$) ellipse (0.4 and 0.7);
	\draw ($(a3)+(0,-0.6)$) ellipse (0.4 and 0.7);
\end{tikzpicture} \right) \propto \frac{x}{x^2-1} \, \LS \left(
\begin{tikzpicture}[baseline={([yshift=-0.1cm]current bounding box.center)}] 
	\node[] (a) at (0,0) {};
	\node[] (a1) at (0.4,0) {};
	\node[] (a2) at (2.4,0) {};
	\node[] (a3) at (2.8,0) {};
	\node[] (b) at (0,-1.2) {};
	\node[] (b1) at (0.4,-1.2) {};
	\node[] (b2) at (2.4,-1.2) {};
	\node[] (b3) at (2.8,-1.2) {};
	\node[] (p1) at ($(a)+(-0.3,0)$) {};
	\node[] (p2) at ($(b)+(-0.3,0)$) {};
	\node[] (p3) at ($(b3)+(0.3,0)$) {};
	\node[] (p4) at ($(a3)+(0.3,0)$) {};
	\draw[line width=0.5mm] (p1.center) -- (0.4,0);
	\draw[line width=0.5mm] (0.4,0) -- (2.6,0);
	\draw[line width=0.5mm] (2.6,0) -- (p4.center);
	\draw[line width=0.5mm] (p2.center) -- (0.4,-1.2);
	\draw[line width=0.5mm] (0.4,-1.2) -- (2.6,-1.2);
	\draw[line width=0.5mm] (2.6,-1.2) -- (p3.center);
	\fill[gray!50] ($(a1)+(0,-0.6)$) ellipse (0.4 and 0.7);
	\draw ($(a1)+(0,-0.6)$) ellipse (0.4 and 0.7);
	\fill[gray!50] ($(a2)+(0,-0.6)$) ellipse (0.4 and 0.7);
	\draw ($(a2)+(0,-0.6)$) ellipse (0.4 and 0.7);
\end{tikzpicture} \right),
\end{equation}
where the blobs indicate any tree- or loop-level exchange, and can also be empty on one side. Therefore, we can reduce the leading singularity of all superclassical integrals containing such graviton propagators to lower-loop cases. Since the lower-loop leading singularities are already computed \cite{Frellesvig:2024zph}, we can drop these superclassical integral topologies, which reduces the number of topologies by half; cf.~table~\ref{tab:reduction_diagrams}.

\begin{figure}[t]
\centering
\subfloat[]{\begin{tikzpicture}[baseline={([yshift=-0.1cm]current bounding box.center)}, scale=0.684]
	\node[] (a) at (0,-0.3) {};
	\node[] (a1) at (1.1,-0.3) {};
	\node[] (a2) at (2.7,-0.3) {};
	\node[] (p1) at (0.69,-0.3) {};
	\node[] (p4) at (3.31,-0.3) {};
	\fill[gray!20] (2,0.2) circle (2.1);
	\fill[white] (2,0.2) circle (1.4);
	\draw[line width=0.15mm] (p4.center) arc (0:180:0.8);
	\draw[line width=0.5mm] (p1.center) -- (p4.center);
\end{tikzpicture}} \qquad
\subfloat[]{\begin{tikzpicture}[baseline={([yshift=-0.1cm]current bounding box.center)}, scale=0.9]
	\node[] (a) at (1.6,0.3) {};
	\node[] (a1) at (2.4,0.3) {};
	\node[] (b) at (2,-1.2) {};
	\node[] (p1) at (1.13,0.3) {};
	\node[] (p4) at (2.87,0.3) {};
	\fill[gray!20] (2,-0.2) circle (1.6);
	\fill[white] (2,-0.2) circle (1);
	\draw[line width=0.15mm] (p1.center) -- (2,-0.4);
	\draw[line width=0.15mm] (a1.center) -- (2,-0.4);
	\draw[line width=0.15mm] (b.center) -- (2,-0.4);
	\draw[line width=0.5mm] (p1.center) -- (p4.center);
\end{tikzpicture}} \qquad \subfloat[]{\begin{tikzpicture}[baseline={([yshift=-0.1cm]current bounding box.center)}, scale=0.9]
	\node[] (a) at (1.45,0.3) {};
	\node[] (a1) at (2.55,0.3) {};
	\node[] (b) at (2,-1.2) {};
	\node[] (p1) at (1.13,0.3) {};
	\node[] (p4) at (2.87,0.3) {};
	\fill[gray!20] (2,-0.2) circle (1.6);
	\fill[white] (2,-0.2) circle (1);
	\draw[line width=0.15mm] (a.center) -- (2,-0.6);
	\draw[line width=0.15mm] (a1.center) -- (2,-0.6);
	\draw[line width=0.15mm] (b.center) -- (2,0.3);
	\draw[line width=0.5mm] (p1.center) -- (p4.center);
\end{tikzpicture}} \qquad \subfloat[]{\begin{tikzpicture}[baseline={([yshift=-0.1cm]current bounding box.center)}, scale=0.9]
	\node[] (a) at (1.45,0.3) {};
	\node[] (a1) at (2.55,0.3) {};
	\node[] (b) at (2,-1.2) {};
	\node[] (p1) at (1.13,0.3) {};
	\node[] (p4) at (2.87,0.3) {};
	\fill[gray!20] (2,-0.2) circle (1.6);
	\fill[white] (2,-0.2) circle (1);
	\draw[line width=0.15mm] (a.center) -- (2,-0.6);
	\draw[line width=0.15mm] (a1.center) -- (2,-0.6);
	\draw[line width=0.15mm] (1.817,0.3) -- (2,-0.6);
	\draw[line width=0.15mm] (2.184,0.3) -- (2,-0.6);
	\draw[line width=0.15mm] (b.center) -- (2,-0.6);
	\draw[line width=0.5mm] (p1.center) -- (p4.center);
\end{tikzpicture}}
\caption{Subgraphs for which the corresponding Feynman integral is reducible to subsectors: (a) One-loop bubble with at least one vertex being cubic; (b) One-loop triangle with at least one cubic vertex at a matter line and cubic graviton self-interaction; (c) Two-loop dangling triangle with cubic matter vertices and a quartic graviton self-interaction; (d) Three-loop dangling triangle with cubic matter vertices and a quintic graviton self-interaction. These drawings should be conceived as being embedded in a bigger graph, which is indicated by the grey zone.}
\label{fig: diagrams_zero_masters}
\end{figure}
Similarly, there exist one-loop IBP relations which guarantee that PM integrals containing certain subgraphs have zero master integrals in their respective sector \cite{Frellesvig:2024zph}. This applies to topologies containing a one-loop bubble subgraph where at least one of its vertices is cubic, see fig.~\ref{fig: diagrams_zero_masters}(a); as well as to those containing a one-loop triangle subgraph where the graviton self-interaction vertex and at least one other vertex are cubic, see fig.~\ref{fig: diagrams_zero_masters}(b). Having zero master integrals means that these integrals are completely reducible to subsectors. As a consequence, they too can be excluded from the analysis, which reduces the number of topologies by one order of magnitude.

For the current four-loop analysis, we have noticed that the triangle reduction appears to have a higher-loop generalization, which further reduces the number of integral topologies by $10\%$. Concretely, integrals containing a subgraph of the type depicted in fig.~\ref{fig: diagrams_zero_masters}(c) and fig.~\ref{fig: diagrams_zero_masters}(d), which following ref.~\cite{Frellesvig:2023bbf} we call multiloop dangling triangles, are likewise reducible to subsectors. While in ref.~\cite{Frellesvig:2024zph} we were able to prove the reduction for fig.~\ref{fig: diagrams_zero_masters}(a) and~fig.~\ref{fig: diagrams_zero_masters}(b) via IBP relations for the subgraph, we have not been able to prove it in full generality for these multiloop dangling triangle subgraphs. We have nevertheless verified up to four loops that all PM integral topologies containing them are indeed reducible to subsectors. It would be interesting to find either a general proof or a counterexample to this multiloop reduction for subgraphs of PM integrals at arbitrarily high loop order.

Lastly, for the purpose of computing the leading singularity, the vertices at the matter lines can be regarded as orderless. Therefore, we can re-order the vertices of certain non-planar integrals and relate their leading singularity to planar counterparts,
\begin{equation}
\label{eq: unraveling_matter_props}
\LS \left( \begin{tikzpicture}[baseline={([yshift=-0.1cm]current bounding box.center)}, scale=0.72]
	\node[] (a) at (0,0.3) {};
	\node[] (a1) at (1,0.3) {};
	\node[] (a2) at (3,0.3) {};
	\node[] (b) at (0,-1) {};
	\node[] (bmod1) at (0.87,-1.03) {};
	\node[] (bmod2) at (1.28,-1.4) {};
	\node[] (bmod3) at (2.72,-1.4) {};
	\node[] (bmod4) at (3.13,-1.03) {};
	\node[] (aend) at (4,0.3) {};
	\node[] (bend) at (4,-1) {};
	\node[] (p1) at (0.69,0.3) {};
	\node[] (p4) at (3.31,0.3) {};
	\fill[gray!20] (2,-0.2) circle (2.1);
	\fill[white] (2,-0.2) circle (1.4);
	\draw[line width=0.15mm] (bmod1.center) -- (a2.center);
	\draw[line width=0.15mm] (bmod2.center) -- (a2.center);
	\node at (1.125,-1)[circle,fill,inner sep=0.6pt]{};
	\node at (1.3,-1)[circle,fill,inner sep=0.6pt]{};
	\node at (1.475,-1)[circle,fill,inner sep=0.6pt]{};
	\draw[line width=0.15mm] (bmod3.center) -- (2.11,-0.8);
	\draw[line width=0.15mm] (1.67,-0.37) -- (a1.center);
	\draw[line width=0.15mm] (bmod4.center) -- (2.37,-0.55);
	\draw[line width=0.15mm] (1.85,-0.23) -- (a1.center);
	\node at (2.525,-1)[circle,fill,inner sep=0.6pt]{};
	\node at (2.7,-1)[circle,fill,inner sep=0.6pt]{};
	\node at (2.875,-1)[circle,fill,inner sep=0.6pt]{};
	\draw[line width=0.5mm] (p1.center) -- (p4.center);
\end{tikzpicture} \right) \propto \LS \left( \begin{tikzpicture}[baseline={([yshift=-0.1cm]current bounding box.center)}, scale=0.72]
	\node[] (a) at (0,0.3) {};
	\node[] (a1) at (1.2,0.3) {};
	\node[] (a2) at (2.8,0.3) {};
	\node[] (b) at (0,-1) {};
	\node[] (bmod1) at (0.87,-1.03) {};
	\node[] (bmod2) at (1.5,-1.51) {};
	\node[] (bmod3) at (2.5,-1.51) {};
	\node[] (bmod4) at (3.13,-1.03) {};
	\node[] (aend) at (4,0.3) {};
	\node[] (bend) at (4,-1) {};
	\node[] (p1) at (0.69,0.3) {};
	\node[] (p4) at (3.31,0.3) {};
	\fill[gray!20] (2,-0.2) circle (2.1);
	\fill[white] (2,-0.2) circle (1.4);
	\draw[line width=0.15mm] (bmod1.center) -- (a1.center);
	\draw[line width=0.15mm] (bmod2.center) -- (a1.center);
	\node at (1,-0.9)[circle,fill,inner sep=0.6pt]{};
	\node at (1.15,-0.9)[circle,fill,inner sep=0.6pt]{};
	\node at (1.3,-0.9)[circle,fill,inner sep=0.6pt]{};
	\draw[line width=0.15mm] (bmod3.center) -- (a2.center);
	\draw[line width=0.15mm] (bmod4.center) -- (a2.center);
	\node at (2.7,-0.9)[circle,fill,inner sep=0.6pt]{};
	\node at (2.85,-0.9)[circle,fill,inner sep=0.6pt]{};
	\node at (3,-0.9)[circle,fill,inner sep=0.6pt]{};
	\draw[line width=0.5mm] (p1.center) -- (p4.center);
\end{tikzpicture} \right),
\end{equation}
where the dots indicate that the vertices can involve three legs or more. We call this operation unraveling, and it significantly reduces the number of independent topologies that need to be considered. While up to three loops we were able to relate the leading singularity of all non-planar integrals to planar counterparts using the unraveling operation~\cite{Frellesvig:2024zph}, a new feature at four loops is the appearance of five truly non-planar integral topologies; see integral topologies $\diagramnumberingsign 41$, $\diagramnumberingsign43$, $\diagramnumberingsign47$, $\diagramnumberingsign48$ and $\diagramnumberingsign55$ in table~\ref{tab: 4-loop results 2SF}.

Overall, we obtain the reduction shown in table~\ref{tab:reduction_diagrams}, which refines and expands the results of ref.~\cite{Frellesvig:2024zph} to four loops. Notice that the final number of integral topologies at three loops is lower than the result in ref.~\cite{Frellesvig:2024zph}, as we identified one further topology which is reducible by the multiloop dangling triangle rule. At four loops, we initially have 16,596 topologies contributing to the classical limit, i.e.\ classical and superclassical ones. Applying the aforementioned simplifying relations, we achieve a drastic reduction in the number of topologies that need to be analyzed, becoming just 70. 

\begin{table}[t]
\begin{center}
\begin{tabular}{lcccc}
\toprule
$L$ & 1 & 2 & 3 & 4 \\[0.1cm] \toprule
Initial number & 2 & 23 & 531 & 16,596 \\ \midrule
Reduction of superclassical integral topologies & 1 & 12 & 271 & 8,335 \\ \midrule
Reduction of one-loop bubbles & 1 & 7 & 100 & 1,854 \\ \midrule
Reduction of one-loop triangles & 1 & 4 & 33 & 476 \\ \midrule
Reduction of multiloop dangling triangles & 1 & 4 & 31 & 422 \\ \midrule
Unraveling of matter propagators & 1 & 4 & 14 & 70 \\[0.1cm] \bottomrule
\end{tabular}
\end{center}
\caption{Number of remaining independent integral topologies after applying each simplifying relation, up to 4 loops. The initial number excludes topologies that do not contribute in the classical limit, reflections, flips of one matter line and integrals that factorize~\cite{Frellesvig:2024zph}.}
\label{tab:reduction_diagrams}
\end{table}

We note that only integrals of the so-called Mondrian type remain after applying the simplifying relations up to four loops, in agreement with what was previously stated by some of the authors in refs.~\cite{Frellesvig:2023bbf,Frellesvig:2024zph}. 

\section{Results: Feynman integral geometries at four loops}
\label{sec:results}

In this section, we present the results for the analysis of Feynman integral geometries contributing to the scattering of black holes at 5PM order, i.e.\ at four loops, thus expanding the three-loop analysis of ref.~\cite{Frellesvig:2024zph}. As discussed in the previous section, the non-trivial geometries present at this order can be classified by analyzing only a subset of 70 different integral topologies, c.f. table~\ref{tab:reduction_diagrams}. In particular, we calculate the leading singularity for one integral of each parity for all of the 70 integral topologies.
 The details of the specific parametrization, loop-by-loop integration sequence used and the result of the Baikov representation are collected in the ancillary files, together with the changes of variables used to calculate each leading singularity.

The results are presented in tables~\ref{tab: 4-loop results 0SF and 1SF} and~\ref{tab: 4-loop results 2SF}, where we indicate higher propagator powers with a dot, and ISPs by multiplication with the corresponding scalar products. In order to specify which loop momenta the ISPs are associated with, we include the label for one representative propagator. Note that for matter propagators only, we use the short-hand notation $2 u_i \cdot k_j \to k_j$ in the labels for compactness. 

The tables are organized with respect to the self-force order of the Feynman integrals.\footnote{Note that, since superclassical integrals can yield different classical subsectors, they can contribute to different self-force orders. In the tables, superclassical integral topologies are included in the lowest self-force order that they enter.} In particular, table~\ref{tab: 4-loop results 0SF and 1SF} includes the integral topologies contributing to 0SF order in rows $\diagramnumberingsign1$ and $\diagramnumberingsign2$, and the 1SF topologies in rows $\diagramnumberingsign3$ to $\diagramnumberingsign36$. The 2SF topologies are gathered in table~\ref{tab: 4-loop results 2SF}, and are numbered from $\diagramnumberingsign37$ to $\diagramnumberingsign70$. For better reference, the first rows in each self-force order gather the only topologies which depend on non-trivial geometries. In particular, this applies to the integral topologies $\diagramnumberingsign3$ to $\diagramnumberingsign6$ and topologies $\diagramnumberingsign37$ to $\diagramnumberingsign40$. The expressions for the polynomials $P_i(\vec{z})$ and $Q_i(\vec{z})$ appearing in their leading singularities can be found in the respective subsections within sec.~\ref{sec:results_CY} and sec.~\ref{sec:results_K3}. For each self-force order, the remaining integral topologies are furthermore ordered based on the number of propagators, from top sectors to subsectors.

For integral topology $\diagramnumberingsign41$, which corresponds to the non-planar 2SF top topology~\cite{Frellesvig:2024zph}, we found it simpler to calculate the result for the leading singularity in six dimensions. This is allowed since the result of the six-dimensional integral can be related to the four-dimensional one by dimension-shift identities~\cite{Tarasov:1996br,Lee:2012te}; thus, the geometry found through its leading singularity is the same. Lastly, the $\varepsilon$-dependence that appears in some of the results below arises when performing the Laurent expansion of the integrand in order to calculate the residue at a double pole.

As we saw in fig.~\ref{fig: non-trivial_diagrams_4loop}, there are in total 8 distinct topologies involving non-trivial geometries at four loops, while the remaining integral topologies have an algebraic leading singularity. We will study these 8 integral topologies in detail in the following subsections. Firstly, integral topologies $\diagramnumberingsign3$ and $\diagramnumberingsign37$ each involve a different three-dimensional Calabi--Yau geometry, as already found in refs.~\cite{Frellesvig:2023bbf,Klemm:2024wtd}. Then, there are 6 integral topologies which contain an integral over a K3 surface. Integral topologies $\diagramnumberingsign4$, $\diagramnumberingsign5$, $\diagramnumberingsign6$, $\diagramnumberingsign38$ and $\diagramnumberingsign39$ depend on the same K3 surface already found at three loops~\cite{Ruf:2021egk,Dlapa:2022wdu,Frellesvig:2024zph}, whereas integral topology $\diagramnumberingsign40$ involves a new K3 surface. Of those, only integral topologies $\diagramnumberingsign4$, $\diagramnumberingsign5$, $\diagramnumberingsign6$ and $\diagramnumberingsign40$ had hitherto been identified in the literature~\cite{Klemm:2024wtd}. 

Let us stress, however, that these 8 integral topologies are not the only ones at four loops which depend on non-trivial geometries. Rather, any integral topology whose leading singularity can be related to the leading singularity of these 8 ones by the relations gathered in sec.~\ref{sec:review_relating_geometries} will also depend on such geometries. For instance, by using the unraveling of matter propagators from eq.~\eqref{eq: unraveling_matter_props} on these 8 topologies, we can easily obtain non-planar variations which also depend on a three-dimensional Calabi--Yau geometry or a K3 surface. Similarly, the dependence on non-trivial geometries can also be inherited from the subsectors through the differential equation that the master integrals satisfy, see sec.~\ref{sec:review_DE_LS}. In particular, undoing the superclassical reduction by adding a rung to the three-loop integral topology that contains a K3 surface~\cite{Ruf:2021egk,Dlapa:2022wdu,Frellesvig:2024zph}, we obtain full agreement with the list of integrals with non-trivial geometries~\cite{Klemm:2024wtd} that occur in the calculation at 1SF order of refs.~\cite{Driesse:2024xad,Bern:2024adl,Driesse:2024feo}.

In the remainder of this section, we study in detail the 8 independent integral topologies which depend on non-trivial geometries, particularly showing the changes \makebox[\linewidth][s]{of variables that are needed to calculate the leading singularity. Moreover, for the}

\begin{table}[ht!]
    \centering
    \caption{Results for the leading singularity of one Feynman integral of each parity for the integral topologies at 0SF order (rows $\diagramnumberingsign1$ and $\diagramnumberingsign2$) and 1SF order (rows $\diagramnumberingsign3$ to $\diagramnumberingsign36$). To specify the ISPs, we include the label for one propagator, where we use the short-hand notation $2 u_i \cdot k_j \to k_j$ for the matter propagators. The polynomials $P_6$ and $P_8$ are specified in eqs.~\eqref{eq: 3-loop K3} and \eqref{eq: P8}, respectively.}
    \label{tab: 4-loop results 0SF and 1SF}
    \vspace{0.1cm}
\begin{tabular}{|C{0.4cm}|L{3.7cm}|C{2.2cm}|L{3.7cm}|C{2.8cm}|}
    \noalign{\global\arrayrulewidth=0.2ex}\hline
        \noalign{\global\arrayrulewidth=0.4pt} 
    & \multicolumn{2}{c|}{Even parity} & \multicolumn{2}{c|}{Odd parity} \\
    \hline
    \# & \multicolumn{1}{c|}{Integral} & LS & \multicolumn{1}{c|}{Integral} & LS \\
\end{tabular}
\end{table} 

\begin{table}[ht!]
    \centering
    \begin{tabular}{|C{0.4cm}|L{3.5cm}|C{2.4cm}|L{4.1cm}|C{2.4cm}|}
    \noalign{\global\arrayrulewidth=0.2ex}\hline
        \noalign{\global\arrayrulewidth=0.4pt} 
    & \multicolumn{2}{c|}{Even parity} & \multicolumn{2}{c|}{Odd parity} \\
    \hline
    \# & \multicolumn{1}{c|}{Integral} & LS & \multicolumn{1}{c|}{Integral} & LS \\
\end{tabular}
\end{table} 

\begin{table}[ht!]
    \centering
    \begin{tabular}{|C{0.4cm}|L{3.5cm}|C{2.4cm}|L{4.1cm}|C{2.4cm}|}
    \noalign{\global\arrayrulewidth=0.2ex}\hline
        \noalign{\global\arrayrulewidth=0.4pt} 
    & \multicolumn{2}{c|}{Even parity} & \multicolumn{2}{c|}{Odd parity} \\
    \hline
    \# & \multicolumn{1}{c|}{Integral} & LS & \multicolumn{1}{c|}{Integral} & LS \\ \noalign{\global\arrayrulewidth=0.2ex}\hline
        \noalign{\global\arrayrulewidth=0.4pt} 
        
& & & & \\
\end{tabular}
\end{table}  

\begin{table}[ht!]
    \centering
    \caption{Results for the leading singularity of one Feynman integral of each parity for the 2SF integral topologies. To specify the ISPs, we include the label for one propagator, where we use the short-hand notation $2 u_i \cdot k_j \to k_j$ for the matter propagators. A superscript $(6D)$ indicates that the leading singularity has been computed in six dimensions.  The polynomials $P_6$, $Q_6$ and $Q_8$ are specified in eqs.~\eqref{eq: 3-loop K3}, \eqref{eq: Q6} and \eqref{eq: Q8}, respectively.}
    \label{tab: 4-loop results 2SF}
    \vspace{0.1cm}
\begin{tabular}{|C{0.4cm}|L{3.85cm}|C{2.4cm}|L{4.1cm}|C{2.05cm}|}
    \noalign{\global\arrayrulewidth=0.2ex}\hline
        \noalign{\global\arrayrulewidth=0.4pt} 
    & \multicolumn{2}{c|}{Even parity} & \multicolumn{2}{c|}{Odd parity} \\
    \hline
    \# & \multicolumn{1}{c|}{Integral} & LS & \multicolumn{1}{c|}{Integral} & LS \\
\end{tabular}
\end{table}  

\begin{table}[ht!]
    \centering
    \begin{tabular}{|C{0.4cm}|L{3.85cm}|C{2.4cm}|L{4.1cm}|C{2.05cm}|}
    \noalign{\global\arrayrulewidth=0.2ex}\hline
        \noalign{\global\arrayrulewidth=0.4pt} 
    & \multicolumn{2}{c|}{Even parity} & \multicolumn{2}{c|}{Odd parity} \\
    \hline
    \# & \multicolumn{1}{c|}{Integral} & LS & \multicolumn{1}{c|}{Integral} & LS \\
\end{tabular}
\end{table}

\begin{table}[ht!]
    \centering
    \begin{tabular}{|C{0.4cm}|L{3.5cm}|C{2.4cm}|L{4.1cm}|C{2.4cm}|}
    \noalign{\global\arrayrulewidth=0.2ex}\hline
        \noalign{\global\arrayrulewidth=0.4pt} 
    & \multicolumn{2}{c|}{Even parity} & \multicolumn{2}{c|}{Odd parity} \\
    \hline
    \# & \multicolumn{1}{c|}{Integral} & LS & \multicolumn{1}{c|}{Integral} & LS \\
\end{tabular}
\end{table}

\clearpage

\noindent two integral topologies involving a Calabi--Yau geometry in one parity, we also prove that they have an algebraic leading singularity in the opposite parity, showcasing how such a small variation can lead to drastically different Feynman integral geometries. In sec.~\ref{sec:results_CY}, we first study the 2 integral topologies which depend on a three-dimensional Calabi--Yau geometry, while we turn to the 6 integral topologies involving a K3 surface in sec.~\ref{sec:results_K3}.

\subsection{Non-trivial geometries I: three-dimensional Calabi--Yau geometries}
\label{sec:results_CY}

In this subsection, we analyze in detail the 2 integral topologies which depend on three-dimensional Calabi--Yau geometries, listed as integral topologies $\diagramnumberingsign3$ and $\diagramnumberingsign37$ in tables~\ref{tab: 4-loop results 0SF and 1SF} and~\ref{tab: 4-loop results 2SF}, respectively. In particular, we show that they depend on a CY geometry for one parity, while becoming polylogarithmic for the opposite parity.

We begin in sec.~\ref{sec:CY1} by studying the 1SF integral topology, which was already identified to be a CY integral through differential equations in ref.~\cite{Klemm:2024wtd}. In sec.~\ref{sec:CY2}, we then turn to the 2SF integral topology, which was already analyzed via leading singularities in ref.~\cite{Frellesvig:2023bbf}. 

In the following, we will rescale the even- and odd-parity integration variables as $z_i \to q^2\, z_i$ and $z_i \to |q| \, z_i$, respectively, in order to transfer the dependence on $|q|$ to the prefactor. This way, the final expressions will become more compact.

\subsubsection{A three-dimensional Calabi--Yau geometry at 1SF order}
\label{sec:CY1}

\begin{figure}[t]
\centering
\begin{tikzpicture}[baseline=(current bounding box.center), scale=0.9] 
	\node[] (a) at (-2,0) {};
	\node[] (a1) at (0,0) {};
	\node[] (a2) at (4,0) {};
	\node[] (a3) at (6,0) {};
	\node[] (b) at (2,-2) {};
	\node[] (c) at (0,-4) {};
	\node[] (c1) at (4,-4) {};
	\node[] (p1) at ($(a)+(-0.5,0)$) {};
	\node[] (p2) at ($(c)+(-2.5,0)$) {};
	\node[] (p3) at ($(c1)+(2.5,0)$) {};
	\node[] (p4) at ($(a3)+(0.5,0)$) {};
	\draw[line width=0.15mm, postaction={decorate}] (b.center) -- node[sloped, allow upside down, label={[xshift=0.1cm, yshift=0.4cm]$k_1$}] {\midarrow} (a1.center);
	\draw[line width=0.15mm, postaction={decorate}] (a2.center) -- node[sloped, allow upside down, label={[xshift=-0.1cm, yshift=0.35cm]$k_1{-}k_3$}] {\midarrow} (b.center);
	\draw[line width=0.15mm, postaction={decorate}] (a3.center) -- node[sloped, allow upside down, label={[xshift=-0.1cm, yshift=0.4cm]$k_3{+}k_4$}] {\midarrow} (c1.center);
	\draw[line width=0.15mm, postaction={decorate}] (c.center) -- node[sloped, allow upside down, label={[xshift=0.15cm, yshift=0.4cm]$k_4{+}q$}] {\midarrow} (a.center);
	\draw[line width=0.15mm, postaction={decorate}] (c1.center) -- node[sloped, allow upside down, label={[xshift=1.55cm, yshift=0.65cm]$k_2{+}k_3$}] {\midarrow} (b.center);
	\draw[line width=0.15mm, postaction={decorate}] (b.center) -- node[sloped, allow upside down, label={[xshift=-0.8cm, yshift=0.65cm]$k_2$}] {\midarrow} (c.center);
	\draw[line width=0.5mm, postaction={decorate}] (a.center) -- node[sloped, allow upside down, label={[xshift=0cm, yshift=-0.2cm]$2u_1 {\cdot} k_4$}] {\midarrow} (a1.center);
	\draw[line width=0.5mm, postaction={decorate}] (a1.center) -- node[sloped, allow upside down, label={[xshift=0cm, yshift=-0.2cm]$2u_1 {\cdot} (k_1{+}k_4)$}] {\midarrow} (a2.center);
	\draw[line width=0.5mm, postaction={decorate}] (a2.center) -- node[sloped, allow upside down, label={[xshift=0cm, yshift=-0.2cm]$2u_1 {\cdot} (k_3{+}k_4)$}] {\midarrow} (a3.center);
	\draw[line width=0.5mm, postaction={decorate}] (c.center) -- node[sloped, allow upside down, label={[xshift=0cm, yshift=-1cm]$2u_2 {\cdot} (k_2{-}k_4)$}] {\midarrow} (c1.center);
	\draw[line width=0.5mm] (p1.center) -- (p4.center);
	\draw[line width=0.5mm] (p2.center) -- (p3.center);
\end{tikzpicture}
\caption{Parametrization of the loop momenta for the 1SF integral topology $\diagramnumberingsign3$ of table~\ref{tab: 4-loop results 0SF and 1SF}, which depends on a three-dimensional CY geometry in the odd-parity sector.}
\label{fig: diag_CY_1}
\end{figure}
Let us first study the 1SF integral topology $\diagramnumberingsign3$ of table~\ref{tab: 4-loop results 0SF and 1SF}, which we parametrize as shown in fig.~\ref{fig: diag_CY_1}. For the even-parity sector, we can consider the scalar integral under the integration order $\{ k_1,k_2,k_3,k_4 \}$. Introducing the ISPs $z_{11}=k_3^2$, $z_{12}=k_4^2$, $z_{13}=2u_2 \cdot k_3$ and $z_{14}=2u_2 \cdot k_4$, we obtain
{\allowdisplaybreaks
\begin{align}
\LS \left( \, \begin{tikzpicture}[baseline={([yshift=-0.1cm]current bounding box.center)}] 
    \node[] (a) at (-0.5,0) {};
	\node[] (a1) at (0,0) {};
    \node[] (a2) at (1,0) {};
    \node[] (a3) at (1.5,0) {};
	\node[] (b) at (0.5,-0.5) {};
	\node[] (c) at (0,-1) {};
    \node[] (c1) at (1,-1) {};
    \draw[line width=0.15mm] (c.center) --  (a.center);
	\draw[line width=0.15mm] (a1.center) --  (b.center);
	\draw[line width=0.15mm] (c.center) --  (b.center);
	\draw[line width=0.15mm] (b.center) --  (c1.center);
	\draw[line width=0.5mm] (a.center) --  (a3.center);
	\draw[line width=0.15mm] (b.center) --  (a2.center);
	\draw[line width=0.5mm] (c.center) --  (c1.center);
	\draw[line width=0.5mm] (-0.65,0) -- (a.center);
	\draw[line width=0.5mm] (a3.center) -- (1.65,0);
	\draw[line width=0.5mm] (-0.65,-1) -- (c.center);
	\draw[line width=0.15mm] (c1.center) --  (a3.center);
	\draw[line width=0.5mm] (c1.center) -- (1.65,-1);
    \end{tikzpicture} \, \right) \propto & \, \LS \biggl( \int \frac{x^2 \, d z_{11} \cdots d z_{14}}{\sqrt{z_{11}} \sqrt{4z_{11}-z_{13}^2} \sqrt{(x^2-1)^2(z_{12}-1)^2-4x^2z_{14}^2}} \nonumber \\
   & \, \times \Big[(x^2-1)^2 z_{11}^2 + z_{12} \Big( (x^2-1)^2 z_{12} - 4 x^2 z_{13} (z_{13} + z_{14}) \Big) \nonumber \\
   & \,  - 
 2 z_{11} \Big( (x^2-1)^2 z_{12} + 2 x^2 z_{14} (z_{13} + z_{14}) \Big) \Big]^{-1/2}  \biggr)\,.
\end{align}
}
At this point, we can use the change of variables in eq.~\eqref{eq: change_of_variables_(z-r1)(z-r2)} from $z_{11}$ to $t_{11}$ to simultaneously rationalize the first two square roots, as well as from $z_{12}$ to $t_{12}$ to rationalize the third one. This generates a single square root in the denominator that is only quadratic in $z_{14}$. Rationalizing this square root with the same change of variables exposes a simple pole at $t_{14}=0$. Taking the residue, we obtain once again a square root in the denominator which is quadratic in $z_{13}$. Rationalizing it with respect $z_{13}$ exposes a simple pole in all remaining integration variables, after which we can easily find
\begin{equation}
\LS \left( \, \begin{tikzpicture}[baseline={([yshift=-0.1cm]current bounding box.center)}] 
    \node[] (a) at (-0.5,0) {};
	\node[] (a1) at (0,0) {};
    \node[] (a2) at (1,0) {};
    \node[] (a3) at (1.5,0) {};
	\node[] (b) at (0.5,-0.5) {};
	\node[] (c) at (0,-1) {};
    \node[] (c1) at (1,-1) {};
    \draw[line width=0.15mm] (c.center) --  (a.center);
	\draw[line width=0.15mm] (a1.center) --  (b.center);
	\draw[line width=0.15mm] (c.center) --  (b.center);
	\draw[line width=0.15mm] (b.center) --  (c1.center);
	\draw[line width=0.5mm] (a.center) --  (a3.center);
	\draw[line width=0.15mm] (b.center) --  (a2.center);
	\draw[line width=0.5mm] (c.center) --  (c1.center);
	\draw[line width=0.5mm] (-0.65,0) -- (a.center);
	\draw[line width=0.5mm] (a3.center) -- (1.65,0);
	\draw[line width=0.5mm] (-0.65,-1) -- (c.center);
	\draw[line width=0.15mm] (c1.center) --  (a3.center);
	\draw[line width=0.5mm] (c1.center) -- (1.65,-1);
    \end{tikzpicture} \, \right) \propto \frac{x}{x^2-1} \, \LS \left( \int \frac{dt_{11} \, dt_{12} \,  dt_{13}}{\sqrt{t_{11}} (t_{11}-1) t_{12} t_{13}} \right) \propto \frac{x}{x^2-1} \,.
\end{equation}
Since the result is algebraic, we conclude that the even-parity contribution for this integral topology is polylogarithmic at the maximal cut, which is in agreement with explicit computations~\cite{Klemm:2024wtd,Driesse:2024xad,Bern:2024adl}.

By contrast, for the odd-parity sector we can for instance add a dot on one matter propagator, which yields
\begin{align}
& \, \LS \left( \, \begin{tikzpicture}[baseline={([yshift=-0.1cm]current bounding box.center)}] 
    \node[] (a) at (-0.5,0) {};
	\node[] (a1) at (0,0) {};
    \node[] (a2) at (1,0) {};
    \node[] (a3) at (1.5,0) {};
	\node[] (b) at (0.5,-0.5) {};
	\node[] (c) at (0,-1) {};
    \node[] (c1) at (1,-1) {};
    \draw[line width=0.15mm] (c.center) --  (a.center);
	\draw[line width=0.15mm] (a1.center) --  (b.center);
	\draw[line width=0.15mm] (c.center) --  (b.center);
	\draw[line width=0.15mm] (b.center) --  (c1.center);
	\draw[line width=0.5mm] (a.center) --  (a3.center);
	\draw[line width=0.15mm] (b.center) --  (a2.center);
	\draw[line width=0.5mm] (c.center) --  (c1.center);
	\draw[line width=0.5mm] (-0.65,0) -- (a.center);
	\draw[line width=0.5mm] (a3.center) -- (1.65,0);
	\draw[line width=0.5mm] (-0.65,-1) -- (c.center);
	\draw[line width=0.15mm] (c1.center) --  (a3.center);
	\draw[line width=0.5mm] (c1.center) -- (1.65,-1);
	\node at (0.5,-1) [circle,fill,inner sep=1.5pt]{};
    \end{tikzpicture} \, \right) \propto \frac{1}{|q|} \, \LS \biggl( \int \frac{x^2 \, d z_{11} \cdots d z_{14}}{\sqrt{z_{11}} \sqrt{4z_{11}-z_{13}^2} \sqrt{(x^2-1)^2(z_{12}-1)^2-4x^2z_{14}^2}} \nonumber \\
   & \quad \, \times  \frac{\varepsilon \left( z_{13}+2z_{14} \right)}{z_{11} + z_{14} (z_{13} + z_{14})} \,  \Big[(x^2-1)^2 z_{11}^2 + z_{12} \Big( (x^2-1)^2 z_{12} - 4 x^2 z_{13} (z_{13} + z_{14}) \Big) \nonumber \\
   & \quad \,  - 
 2 z_{11} \Big( (x^2-1)^2 z_{12} + 2 x^2 z_{14} (z_{13} + z_{14}) \Big) \Big]^{-1/2} \biggr)\,.
\end{align}
As can be seen, doubling the matter propagator generates an extra factor that depends on $z_{11}$, $z_{13}$ and $z_{14}$. This factor spoils the change of variables used for the even-parity sector, as it would now generate further terms which do not allow for a residue at the end. Instead, we can rationalize the first square root with $z_{11}=t_{11}^2$, and afterwards use eq.~\eqref{eq: change_of_variables_(z-r1)(z-r2)} to rationalize the second and third square roots with respect to $z_{13}$ and $z_{12}$, respectively. This generates a single square root which is still quadratic in $z_{14}$, thus allowing for a rationalization and subsequent residue. After rescaling $t_{12} \to t_{12}/t_{11}$ and relabeling $\{ t_{11}, t_{12}, t_{13} \} \to \{ t_1, t_2, t_3 \}$, we finally obtain
\begin{equation}
\LS \left( \, \begin{tikzpicture}[baseline={([yshift=-0.1cm]current bounding box.center)}] 
    \node[] (a) at (-0.5,0) {};
	\node[] (a1) at (0,0) {};
    \node[] (a2) at (1,0) {};
    \node[] (a3) at (1.5,0) {};
	\node[] (b) at (0.5,-0.5) {};
	\node[] (c) at (0,-1) {};
    \node[] (c1) at (1,-1) {};
    \draw[line width=0.15mm] (c.center) --  (a.center);
	\draw[line width=0.15mm] (a1.center) --  (b.center);
	\draw[line width=0.15mm] (c.center) --  (b.center);
	\draw[line width=0.15mm] (b.center) --  (c1.center);
	\draw[line width=0.5mm] (a.center) --  (a3.center);
	\draw[line width=0.15mm] (b.center) --  (a2.center);
	\draw[line width=0.5mm] (c.center) --  (c1.center);
	\draw[line width=0.5mm] (-0.65,0) -- (a.center);
	\draw[line width=0.5mm] (a3.center) -- (1.65,0);
	\draw[line width=0.5mm] (-0.65,-1) -- (c.center);
	\draw[line width=0.15mm] (c1.center) --  (a3.center);
	\draw[line width=0.5mm] (c1.center) -- (1.65,-1);
	\node at (0.5,-1) [circle,fill,inner sep=1.5pt]{};
    \end{tikzpicture} \, \right) \propto \frac{\varepsilon\,x^2}{|q|\sqrt{x^2{-}1}} \int \frac{d t_1 d t_2 d t_3}{\sqrt{P_8(t_1,t_2,t_3)}}\,,
\end{equation}
where
\begin{align}
P_8(t_1,t_2,t_3) =&\,-t_2^2 t_3^2 (t_1^2 - 1)^2 (1 - x^6) + 
 2 t_2 t_3^3 (t_1^2 - 1) (t_1^2 + t_2^2) x (1 + x^4) \nonumber \\
&\, + \left( 4 t_1^2 t_2^2 - t_2^2 t_3^2 (t_1^2 + 3) (t_1^2 - 1) - t_3^4 (t_1^2 + t_2^2)^2 \right) x^2(1-x^2) \nonumber \\
&\, + 
 4 t_2 t_3 (t_1^2 + t_2^2) (t_1^2 + t_3^2) x^3
 \label{eq: P8}
\end{align}
is a polynomial of overall degree 8 in three variables, and is quartic in each. Following the discussion below eq.~\eqref{eq: LS_definition}, we find that this odd-parity integral depends on a three-dimensional Calabi--Yau geometry.\footnote{In fact, after rationalizing and taking the residue in $z_{14}$, we obtain two different leading singularities and three-dimensional Calabi--Yau geometries. However, they are actually equivalent under the change of variables $t_3 \to 1/t_3$. More details can be found in the ancillary files.} This geometry was originally identified via differential equations in ref.~\cite{Klemm:2024wtd}, where an irreducible fourth-order Picard--Fuchs operator was found to annihilate this integral in $D=4$.

\subsubsection{A three-dimensional Calabi--Yau geometry at 2SF order}
\label{sec:CY2}

\begin{figure}[t]
\centering
\begin{tikzpicture}[baseline=(current bounding box.center)] 
	\node[] (a) at (0,0) {};
	\node[] (a1) at (3,0) {};
	\node[] (a2) at (6,0) {};
	\node[] (b) at (0,-2) {};
	\node[] (b1) at (6,-2) {};
	\node[] (c) at (0,-4) {};
	\node[] (c1) at (3,-4) {};
	\node[] (c2) at (6,-4) {};
	\node[] (p1) at ($(a)+(-0.5,0)$) {};
	\node[] (p2) at ($(c)+(-0.5,0)$) {};
	\node[] (p3) at ($(c2)+(0.5,0)$) {};
	\node[] (p4) at ($(a2)+(0.5,0)$) {};
	\draw[line width=0.15mm, postaction={decorate}] (b.center) -- node[sloped, allow upside down, label={[xshift=0.15cm, yshift=0cm]$k_1$}] {\midarrow} (a.center);
	\draw[line width=0.15mm, postaction={decorate}] (a1.center) -- node[sloped, allow upside down, label={[xshift=-0.05cm, yshift=0.45cm]$k_1{-}k_2$}] {\midarrow} (b.center);
	\draw[line width=0.15mm, postaction={decorate}] (b1.center) -- node[sloped, allow upside down, label={[xshift=-1.5cm, yshift=0.65cm]$k_4{-}k_3$}] {\midarrow} (c1.center);
	\draw[line width=0.15mm, postaction={decorate}] (a2.center) -- node[sloped, allow upside down, label={[xshift=-0.15cm, yshift=0cm]$k_2{-}q$}] {\midarrow} (b1.center);
	\draw[line width=0.15mm, postaction={decorate}] (b1.center) -- node[sloped, allow upside down, label={[xshift=0cm, yshift=0.9cm]$k_2{+}k_3$}] {\midarrow} (b.center);
	\draw[line width=0.15mm, postaction={decorate}] (b.center) -- node[sloped, allow upside down, label={[xshift=-0.9cm, yshift=0.05cm]$k_3$}] {\midarrow} (c.center);
	\draw[line width=0.15mm, postaction={decorate}] (c2.center) -- node[sloped, allow upside down, label={[xshift=1.45cm, yshift=0cm]$k_4{+}q$}] {\midarrow} (b1.center);
	\draw[line width=0.5mm, postaction={decorate}] (a.center) -- node[sloped, allow upside down, label={[xshift=0cm, yshift=-0.15cm]$2u_1 {\cdot} k_1$}] {\midarrow} (a1.center);
	\draw[line width=0.5mm, postaction={decorate}] (a1.center) -- node[sloped, allow upside down, label={[xshift=0cm, yshift=-0.15cm]$2u_1 {\cdot} k_2$}] {\midarrow} (a2.center);
	\draw[line width=0.5mm, postaction={decorate}] (c.center) -- node[sloped, allow upside down, label={[xshift=0cm, yshift=-0.9cm]$2u_2 {\cdot} k_3$}] {\midarrow} (c1.center);
	\draw[line width=0.5mm, postaction={decorate}] (c1.center) -- node[sloped, allow upside down, label={[xshift=0cm, yshift=-0.9cm]$2u_2 {\cdot} k_4$}] {\midarrow} (c2.center);
	\draw[line width=0.5mm] (p1.center) -- (p4.center);
	\draw[line width=0.5mm] (p2.center) -- (p3.center);
\end{tikzpicture}
\caption{Parametrization of the loop momenta for the 2SF integral topology $\diagramnumberingsign37$ of table~\ref{tab: 4-loop results 2SF}, which depends on a three-dimensional CY geometry in the even-parity sector.}
\label{fig: diag_CY_2}
\end{figure}

Now, let us study the 2SF integral topology $\diagramnumberingsign37$ of table~\ref{tab: 4-loop results 2SF}, which we parametrize as shown in fig.~\ref{fig: diag_CY_2}. The analysis of the leading singularity of this integral was originally done by some of the present authors in ref.\ \cite{Frellesvig:2023bbf} and is repeated here for the convenience of the reader; see also ref.~\cite{Correia:2025yao}.

For the even-parity sector, we can consider the scalar integral under the integration order $\{ k_1,k_4,k_3,k_2 \}$. Introducing the ISPs $z_{12}=(k_3+q)^2$, $z_{13}=k_2^2$ and $z_{14}=2u_2 \cdot k_2$, we obtain
\begin{align}
\LS \left( \begin{tikzpicture}[baseline={([yshift=-0.1cm]current bounding box.center)}] 
	\node[] (a) at (0,0) {};
	\node[] (a1) at (0.75,0) {};
	\node[] (a2) at (1.5,0) {};
	\node[] (b) at (0,-0.5) {};
	\node[] (b1) at (0.5,-0.5) {};
	\node[] (b2) at (1.5,-0.5) {};
	\node[] (c) at (0,-1) {};
	\node[] (c1) at (0.75,-1) {};
	\node[] (c2) at (1.5,-1) {};
	\node[] (p1) at ($(a)+(-0.2,0)$) {};
	\node[] (p2) at ($(c)+(-0.2,0)$) {};
	\node[] (p3) at ($(c2)+(0.2,0)$) {};
	\node[] (p4) at ($(a2)+(0.2,0)$) {};
	\draw[line width=0.15mm] (c.center) -- (a.center);
	\draw[line width=0.15mm] (b.center) -- (b2.center);
	\draw[line width=0.15mm] (c2.center) -- (a2.center);
	\draw[line width=0.15mm] (b2.center) -- (c1.center);
	\draw[line width=0.15mm] (b.center) -- (a1.center);
	\draw[line width=0.5mm] (p1.center) -- (p4.center);
	\draw[line width=0.5mm] (p2.center) -- (p3.center);
\end{tikzpicture} \right) \propto & \, \frac{1}{q^2} \, \LS \biggl( \int \frac{x \, d z_{12} d z_{13} d z_{14}}{\sqrt{z_{12}} \sqrt{z_{13}} \sqrt{(x^2-1)^2 (z_{13}-1)^2 -4x^2 z_{14}^2}}  \nonumber \\
& \,  \times \, \frac{1}{\sqrt{z_{14}^2 (z_{12}-1)^2 - 4 z_{12} z_{13}(z_{12} + z_{13} - 1)}} \biggr).
\end{align}
Using the change of variables from eq.~\eqref{eq: change_of_variables_(z-r1)(z-r2)} with respect to $z_{14}$ and the third square root, followed by the rescaling $t_{14}\to t_{14}/(\sqrt{z_{12}}\sqrt{z_{13}})$, we can simultaneously rationalize the first three square roots. This generates a single square root in the denominator which is quartic in $z_{12}$, $z_{13}$ and $t_{14}$. After relabeling $\{ z_{12}, z_{13}, t_{14} \}  \to \{ t_1, t_2, t_3 \}$, we obtain
\begin{equation}
\LS \left( \begin{tikzpicture}[baseline={([yshift=-0.1cm]current bounding box.center)}] 
	\node[] (a) at (0,0) {};
	\node[] (a1) at (0.75,0) {};
	\node[] (a2) at (1.5,0) {};
	\node[] (b) at (0,-0.5) {};
	\node[] (b1) at (0.5,-0.5) {};
	\node[] (b2) at (1.5,-0.5) {};
	\node[] (c) at (0,-1) {};
	\node[] (c1) at (0.75,-1) {};
	\node[] (c2) at (1.5,-1) {};
	\node[] (p1) at ($(a)+(-0.2,0)$) {};
	\node[] (p2) at ($(c)+(-0.2,0)$) {};
	\node[] (p3) at ($(c2)+(0.2,0)$) {};
	\node[] (p4) at ($(a2)+(0.2,0)$) {};
	\draw[line width=0.15mm] (c.center) -- (a.center);
	\draw[line width=0.15mm] (b.center) -- (b2.center);
	\draw[line width=0.15mm] (c2.center) -- (a2.center);
	\draw[line width=0.15mm] (b2.center) -- (c1.center);
	\draw[line width=0.15mm] (b.center) -- (a1.center);
	\draw[line width=0.5mm] (p1.center) -- (p4.center);
	\draw[line width=0.5mm] (p2.center) -- (p3.center);
\end{tikzpicture} \right) \propto \frac{x}{q^2} \int \frac{d t_1 d t_2 d t_3}{\sqrt{Q_8(t_1,t_2,t_3)}}\,,
\end{equation}
where
\begin{align}
\label{eq: Q8}
 Q_8(t_1,t_2,t_3) = (t_1-1)^2(t_2-1)^2(t_1t_2+t_3^2)^2(1-x^2)^2-64x^2 t_1^2t_2^2t_3^2(t_1+t_2-1)\,
\end{align}
is a polynomial of overall degree 8. Similarly to sec.~\ref{sec:CY1}, we find that this even-parity integral depends on a three-dimensional Calabi--Yau geometry, as was first identified using leading singularities in ref.~\cite{Frellesvig:2023bbf}. In the same article, it was shown that from the perspective of differential equations one obtains an irreducible Picard--Fuchs operator of order 4 in $D=4$ dimensions, corroborating this result. A way to further characterize this Calabi--Yau operator is by computing the so-called instanton numbers~\cite{Almkvist:2021}; see ref.~\cite{Frellesvig:2024rea} for a discussion in the context of PM integrals. Notably, the instanton numbers associated with this operator differ from those of the 1SF operator discussed in the previous section, demonstrating that the two three-dimensional Calabi--Yau geometries are genuinely different~\cite{Klemm:2024wtd,Frellesvig:2024rea}. Furthermore, in ref.~\cite{Frellesvig:2024rea}, some of the present authors and collaborators derived an $\varepsilon$-factorized differential equation for this even-parity sector. Since the differential equation in $D=4-2\varepsilon$ dimensions exhibits apparent singularities, this required an extension of the techniques developed in refs.~\cite{Pogel:2022ken,Pogel:2022vat}.

Let us then consider the odd-parity sector, which turns out to be more easily evaluated using the integration order $\{ k_1,k_2,k_3,k_4\}$ with the same parametrization of loop momenta as in fig.~\ref{fig: diag_CY_2}. Following eq.~\eqref{eq: n_ISP}, this parametrization requires $\nISP=5$ which we choose to be $z_{12}=k_2^2$, $z_{13}=(k_3-q)^2$, $z_{14}=k_4^2$, $z_{15}=2u_1 \cdot k_3$ and $z_{16}=2u_1 \cdot k_4$. In this case, we can pick an odd-parity integral by multiplying the even-parity integrand by a factor of the odd-parity ISP $z_{15}$, to obtain
\begin{align}
&\LS \left( \begin{tikzpicture}[baseline={([yshift=-0.1cm]current bounding box.center)}] 
	\node[] (a) at (0,0) {};
	\node[] (a1) at (0.75,0) {};
	\node[] (a2) at (1.5,0) {};
	\node[] (b) at (0,-0.5) {};
	\node[] (b1) at (0.5,-0.5) {};
	\node[] (b2) at (1.5,-0.5) {};
	\node[] (c) at (0,-1) {};
	\node[] (c1) at (0.75,-1) {};
	\node[] (c2) at (1.5,-1) {};
	\node[] (p1) at ($(a)+(-0.2,0)$) {};
	\node[] (p2) at ($(c)+(-0.2,0)$) {};
	\node[] (p3) at ($(c2)+(0.2,0)$) {};
	\node[] (p4) at ($(a2)+(0.2,0)$) {};
    \draw[line width=0.15mm] (a.center) -- (b.center);
	\draw[line width=0.15mm] (c2.center) -- (a2.center);
	\draw[line width=0.15mm] (b.center) -- (b2.center);
	\draw[line width=0.15mm] (c2.center) -- (a2.center);
	\draw[line width=0.15mm] (b2.center) -- (c1.center);
    \draw[line width=0.15mm] (b.center) -- (c.center);
	\draw[line width=0.15mm] (b.center) -- (a1.center);
	\draw[line width=0.5mm] (p1.center) -- (p4.center);
	\draw[line width=0.5mm] (p2.center) -- (p3.center);
\end{tikzpicture} {\times} \, 2u_1 {\cdot} k_3 \right) \nonumber \\
& \, \propto \frac{1}{|q|} \, \LS \biggl( \int \frac{\varepsilon \, x^2 \, z_{15}\, d z_{12} \, \cdots \, d z_{16}}{\sqrt{z_{12}} \sqrt{(z_{12}-1)^2 z_{15}^2 + 4z_{12}(z_{13}-2)(1+z_{12}-z_{13})} \ P_4(z_{13},\dots,z_{16}) } \biggr),
\end{align}
where $P_4(z_{13},\dots,z_{16})$ is a polynomial of overall degree 4, but that is only quadratic in $z_{16}$. Therefore, we can immediately compute the residue at either of the simple poles in the integration variable $z_{16}$, leading to
\begin{align}
& \, \LS \left( \begin{tikzpicture}[baseline={([yshift=-0.1cm]current bounding box.center)}] 
	\node[] (a) at (0,0) {};
	\node[] (a1) at (0.75,0) {};
	\node[] (a2) at (1.5,0) {};
	\node[] (b) at (0,-0.5) {};
	\node[] (b1) at (0.5,-0.5) {};
	\node[] (b2) at (1.5,-0.5) {};
	\node[] (c) at (0,-1) {};
	\node[] (c1) at (0.75,-1) {};
	\node[] (c2) at (1.5,-1) {};
	\node[] (p1) at ($(a)+(-0.2,0)$) {};
	\node[] (p2) at ($(c)+(-0.2,0)$) {};
	\node[] (p3) at ($(c2)+(0.2,0)$) {};
	\node[] (p4) at ($(a2)+(0.2,0)$) {};
    \draw[line width=0.15mm] (a.center) -- (b.center);
	\draw[line width=0.15mm] (c2.center) -- (a2.center);
	\draw[line width=0.15mm] (b.center) -- (b2.center);
	\draw[line width=0.15mm] (c2.center) -- (a2.center);
	\draw[line width=0.15mm] (b2.center) -- (c1.center);
    \draw[line width=0.15mm] (b.center) -- (c.center);
	\draw[line width=0.15mm] (b.center) -- (a1.center);
	\draw[line width=0.5mm] (p1.center) -- (p4.center);
	\draw[line width=0.5mm] (p2.center) -- (p3.center);
\end{tikzpicture} {\times} \, 2u_1 {\cdot} k_3 \! \right) \! \propto \frac{1}{|q|} \, \LS  \biggl( \int \! \frac{\varepsilon \, x \, z_{15}\, d z_{12} \, \cdots \, d z_{15}}{\sqrt{(z_{12}-1)^2 z_{15}^2 + 4z_{12}(z_{13}-2)(1{+}z_{12}{-}z_{13})} }  \nonumber \\[0.1cm]
& \, \quad   \times \frac{1}{\sqrt{z_{12}} \sqrt{z_{13}-2} \sqrt{z_{14} (1+z_{14} - z_{13})} \sqrt{(x^2-1)^2 (z_{13}-1)^2 -4x^2 z_{15}^2}} \biggr).
\end{align}
At this point, we can introduce the change of variables in eq.~\eqref{eq: change_of_variables_(z-r1)(z-r2)} from $z_{14}$ to $t_{14}$ to rationalize the fourth square root, as well as transform $z_{15} = \sqrt{z'_{15}}$, which removes the factor in the numerator, since the remaining square roots contain only even powers of $z_{15}$. Afterwards, we can use again the change of variables in eq.~\eqref{eq: change_of_variables_(z-r1)(z-r2)} from $z'_{15}$ to $t_{15}$, in order to rationalize the first and fifth square roots simultaneously. After these transformations, we are left with the result
\begin{align}
\LS \left( \begin{tikzpicture}[baseline={([yshift=-0.1cm]current bounding box.center)}] 
	\node[] (a) at (0,0) {};
	\node[] (a1) at (0.75,0) {};
	\node[] (a2) at (1.5,0) {};
	\node[] (b) at (0,-0.5) {};
	\node[] (b1) at (0.5,-0.5) {};
	\node[] (b2) at (1.5,-0.5) {};
	\node[] (c) at (0,-1) {};
	\node[] (c1) at (0.75,-1) {};
	\node[] (c2) at (1.5,-1) {};
	\node[] (p1) at ($(a)+(-0.2,0)$) {};
	\node[] (p2) at ($(c)+(-0.2,0)$) {};
	\node[] (p3) at ($(c2)+(0.2,0)$) {};
	\node[] (p4) at ($(a2)+(0.2,0)$) {};
    \draw[line width=0.15mm] (a.center) -- (b.center);
	\draw[line width=0.15mm] (c2.center) -- (a2.center);
	\draw[line width=0.15mm] (b.center) -- (b2.center);
	\draw[line width=0.15mm] (c2.center) -- (a2.center);
	\draw[line width=0.15mm] (b2.center) -- (c1.center);
    \draw[line width=0.15mm] (b.center) -- (c.center);
	\draw[line width=0.15mm] (b.center) -- (a1.center);
	\draw[line width=0.5mm] (p1.center) -- (p4.center);
	\draw[line width=0.5mm] (p2.center) -- (p3.center);
\end{tikzpicture} {\times} \, 2u_1 {\cdot} k_3 \right) \propto & \, \frac{\varepsilon}{|q|} \, \LS \left( \int \! \frac{dz_{12} dz_{13} dt_{14} dt_{15}}{t_{14} t_{15} (z_{12}-1) \sqrt{z_{12}} \sqrt{z_{13}-2}} \right) \nonumber \\
\propto & \, \frac{\varepsilon}{|q|} \, \LS \left( \int \frac{d z_{13}}{\sqrt{z_{13}-2}} \right),
\end{align}
where we computed the residue at the simple poles in the last step. As can be seen, we cannot expose any further simple poles in $z_{13}$ via a coordinate transformation. Instead, we follow the discussion in sec.~\ref{sec:review_DE_LS} and integrate the leading singularity over a contour specified by the zeros of a maximally-cut Baikov polynomial. Ultimately, this yields an algebraic leading singularity for the odd-parity sector,
\begin{align}
\LS \left( \begin{tikzpicture}[baseline={([yshift=-0.1cm]current bounding box.center)}] 
	\node[] (a) at (0,0) {};
	\node[] (a1) at (0.75,0) {};
	\node[] (a2) at (1.5,0) {};
	\node[] (b) at (0,-0.5) {};
	\node[] (b1) at (0.5,-0.5) {};
	\node[] (b2) at (1.5,-0.5) {};
	\node[] (c) at (0,-1) {};
	\node[] (c1) at (0.75,-1) {};
	\node[] (c2) at (1.5,-1) {};
	\node[] (p1) at ($(a)+(-0.2,0)$) {};
	\node[] (p2) at ($(c)+(-0.2,0)$) {};
	\node[] (p3) at ($(c2)+(0.2,0)$) {};
	\node[] (p4) at ($(a2)+(0.2,0)$) {};
    \draw[line width=0.15mm] (a.center) -- (b.center);
	\draw[line width=0.15mm] (c2.center) -- (a2.center);
	\draw[line width=0.15mm] (b.center) -- (b2.center);
	\draw[line width=0.15mm] (c2.center) -- (a2.center);
	\draw[line width=0.15mm] (b2.center) -- (c1.center);
    \draw[line width=0.15mm] (b.center) -- (c.center);
	\draw[line width=0.15mm] (b.center) -- (a1.center);
	\draw[line width=0.5mm] (p1.center) -- (p4.center);
	\draw[line width=0.5mm] (p2.center) -- (p3.center);
\end{tikzpicture} {\times} \, 2u_1 {\cdot} k_3 \right) \propto \frac{\varepsilon}{|q|} \int_{1}^{2} \! \frac{d z_{13}}{\sqrt{z_{13}-2}} \propto \frac{\varepsilon}{|q|}\,.
\end{align}

As in the previous subsection, the leading singularity of this integral is thus 
related to a period integral over a three-dimensional Calabi--Yau geometry in one parity sector, while the opposite parity has an algebraic leading singularity and is thus polylogarithmic on the maximal cut.

\subsection{Non-trivial geometries II: K3 surfaces}
\label{sec:results_K3}

In this subsection, we analyze in detail the 6 integral topologies which depend on a two-dimensional K3 surface, listed as numbers $\diagramnumberingsign4$, $\diagramnumberingsign5$, $\diagramnumberingsign6$, $\diagramnumberingsign38$, $\diagramnumberingsign39$ and $\diagramnumberingsign40$ in tables~\ref{tab: 4-loop results 0SF and 1SF} and~\ref{tab: 4-loop results 2SF}. Analogously to sec.~\ref{sec:results_CY}, these integral topologies all depend on a K3 surface for one parity, while having an algebraic leading singularity in the opposite parity. Having exemplified this drastic reduction in the complexity of the associated geometry for the Calabi--Yau integrals in the previous subsection, here we will only focus on the parity sector where the K3 surface arises for each topology. For a full derivation of the algebraic leading singularity in the opposite parity, we refer to the ancillary files.

We begin in sec.~\ref{sec:K31} by studying the first 5 integral topologies, which are all related to the same K3 surface that already appeared at three loops~\cite{Ruf:2021egk,Dlapa:2022wdu,Frellesvig:2024zph}. Two of these topologies (38 and 39) are novel, while the remaining were found to depend on a K3 surface through differential equations in ref.~\cite{Klemm:2024wtd}. In sec.~\ref{sec:K32}, we then turn to the last integral topology, first identified to contain a different K3 surface through differential equations also in ref.~\cite{Klemm:2024wtd}. As in the previous subsection, we will appropriately rescale the integration variables to transfer the dependence on $|q|$ to the prefactor and simplify the expressions.

\subsubsection{The 3-loop K3 surface at 1SF and 2SF orders}
\label{sec:K31}

Let us first analyze the 5 integral topologies that depend on the same K3 surface that already appeared at 3-loop order~\cite{Ruf:2021egk,Dlapa:2022wdu,Frellesvig:2024zph}. The integral topologies in question are $\diagramnumberingsign4$, $\diagramnumberingsign5$, $\diagramnumberingsign6$, $\diagramnumberingsign38$ and $\diagramnumberingsign39$ in tables~\ref{tab: 4-loop results 0SF and 1SF} and~\ref{tab: 4-loop results 2SF}. Unless otherwise stated, we shall henceforth assume the integration order $\{ k_1,k_2,k_3,k_4 \}$.

\begin{figure}[t]
\centering
\subfloat[]{\begin{tikzpicture}[baseline=(current bounding box.center), scale=0.82] 
	\node[] (a) at (0,0) {};
	\node[] (a1) at (2,0) {};
    \node[] (a2) at (4,0) {};
	\node[] (a3) at (6,0) {};
	\node[] (b) at (0,-2) {};
	\node[] (b1) at (3,-2) {};
	\node[] (c) at (0,-4) {};
	\node[] (c1) at (6,-4) {};
	\node[] (p1) at ($(a)+(-0.5,0)$) {};
	\node[] (p2) at ($(c)+(-0.5,0)$) {};
	\node[] (p3) at ($(c1)+(0.5,0)$) {};
	\node[] (p4) at ($(a3)+(0.5,0)$) {};
	\draw[line width=0.15mm, postaction={decorate}] (b.center) -- node[sloped, allow upside down, label={[xshift=0.15cm, yshift=0cm]$k_2{-}k_4$}] {\midarrow} (a.center);
	\draw[line width=0.15mm, postaction={decorate}] (b1.center) -- node[sloped, allow upside down, label={[xshift=0.15cm, yshift=0.4cm]$k_1{-}k_2$}] {\midarrow} (a1.center);
	\draw[line width=0.15mm, postaction={decorate}] (c1.center) -- node[sloped, allow upside down, label={[xshift=0.05cm, yshift=0.45cm]$k_3$}] {\midarrow} (b1.center);
	\draw[line width=0.15mm, postaction={decorate}] (a2.center) -- node[sloped, allow upside down, label={[xshift=-0.15cm, yshift=0.35cm]$k_1$}] {\midarrow} (b1.center);
	\draw[line width=0.15mm, postaction={decorate}] (b1.center) -- node[sloped, allow upside down, label={[xshift=0cm, yshift=0.15cm]$k_2{+}k_3$}] {\midarrow} (b.center);
	\draw[line width=0.15mm, postaction={decorate}] (b.center) -- node[sloped, allow upside down, label={[xshift=-1.6cm, yshift=0.05cm]$k_3{+}k_4$}] {\midarrow} (c.center);
	\draw[line width=0.15mm, postaction={decorate}] (c1.center) -- node[sloped, allow upside down, label={[xshift=1.45cm, yshift=0cm]$k_4{+}q$}] {\midarrow} (a3.center);
	\draw[line width=0.5mm, postaction={decorate}] (a.center) -- node[sloped, allow upside down, label={[xshift=-0.35cm, yshift=-0.15cm]$2u_1 {\cdot} (k_2{-}k_4)$}] {\midarrow} (a1.center);
	\draw[line width=0.5mm, postaction={decorate}] (a1.center) -- node[sloped, allow upside down, label={[xshift=0.35cm, yshift=-0.15cm]$2u_1 {\cdot} (k_1{-}k_4)$}] {\midarrow} (a2.center);
    \draw[line width=0.5mm, postaction={decorate}] (a2.center) -- node[sloped, allow upside down, label={[xshift=-0.05cm, yshift=-0.85cm]${-}2u_1 {\cdot} k_4$}] {\midarrow} (a3.center);
	\draw[line width=0.5mm, postaction={decorate}] (c.center) -- node[sloped, allow upside down, label={[xshift=0cm, yshift=-0.95cm]$2u_2 {\cdot} (k_3{+}k_4)$}] {\midarrow} (c1.center);
	\draw[line width=0.5mm] (p1.center) -- node[] {} (a.center);
	\draw[line width=0.5mm] (a3.center) -- node[] {} (p4.center);
	\draw[line width=0.5mm] (p2.center) -- node[] {} (c.center);
	\draw[line width=0.5mm] (c1.center) -- node[] {} (p3.center);
\end{tikzpicture}} \quad \enspace \subfloat[]{\begin{tikzpicture}[baseline=(current bounding box.center), scale=0.82] 
	\node[] (a) at (0,0) {};
	\node[] (a1) at (2,0) {};
	\node[] (a2) at (4,0) {};
    \node[] (a3) at (6,0) {};
	\node[] (b) at (0,-2) {};
	\node[] (b1) at (4,-2) {};
	\node[] (c) at (0,-4) {};
	\node[] (c1) at (6,-4) {};
	\node[] (p1) at ($(a)+(-0.5,0)$) {};
	\node[] (p2) at ($(c)+(-0.5,0)$) {};
	\node[] (p3) at ($(c1)+(0.5,0)$) {};
	\node[] (p4) at ($(a3)+(0.5,0)$) {};
	\draw[line width=0.15mm, postaction={decorate}] (b.center) -- node[sloped, allow upside down, label={[xshift=0.15cm, yshift=0cm]$k_1$}] {\midarrow} (a.center);
	\draw[line width=0.15mm, postaction={decorate}] (a1.center) -- node[sloped, allow upside down, label={[xshift=-0.05cm, yshift=0.45cm]$k_1{-}k_2$}] {\midarrow} (b.center);
	\draw[line width=0.15mm, postaction={decorate}] (c1.center) -- node[sloped, allow upside down, label={[xshift=0.1cm, yshift=0.45cm]$k_3{-}k_4$}] {\midarrow} (b1.center);
	\draw[line width=0.15mm, postaction={decorate}] (a2.center) -- node[sloped, allow upside down, label={[xshift=-0.2cm, yshift=0cm]$k_2{+}k_4$}] {\midarrow} (b1.center);
	\draw[line width=0.15mm, postaction={decorate}] (b1.center) -- node[sloped, allow upside down, label={[xshift=0cm, yshift=0.15cm]$k_2{+}k_3$}] {\midarrow} (b.center);
	\draw[line width=0.15mm, postaction={decorate}] (b.center) -- node[sloped, allow upside down, label={[xshift=-0.9cm, yshift=0.05cm]$k_3$}] {\midarrow} (c.center);
	\draw[line width=0.15mm, postaction={decorate}] (c1.center) -- node[sloped, allow upside down, label={[xshift=1.45cm, yshift=0cm]$k_4{+}q$}] {\midarrow} (a3.center);
	\draw[line width=0.5mm, postaction={decorate}] (a.center) -- node[sloped, allow upside down, label={[xshift=0cm, yshift=-0.15cm]$2u_1 {\cdot} k_1$}] {\midarrow} (a1.center);
	\draw[line width=0.5mm, postaction={decorate}] (a1.center) -- node[sloped, allow upside down, label={[xshift=0cm, yshift=-0.15cm]$2u_1 {\cdot} k_2$}] {\midarrow} (a2.center);
    \draw[line width=0.5mm, postaction={decorate}] (a2.center) -- node[sloped, allow upside down, label={[xshift=0cm, yshift=-0.15cm]$-2u_1 {\cdot} k_4$}] {\midarrow} (a3.center);
	\draw[line width=0.5mm, postaction={decorate}] (c.center) -- node[sloped, allow upside down, label={[xshift=0cm, yshift=-0.9cm]$2u_2 {\cdot} k_3$}] {\midarrow} (c1.center);
	\draw[line width=0.5mm] (p1.center) -- node[] {} (a.center);
	\draw[line width=0.5mm] (a3.center) -- node[] {} (p4.center);
	\draw[line width=0.5mm] (p2.center) -- node[] {} (c.center);
	\draw[line width=0.5mm] (c1.center) -- node[] {} (p3.center);
\end{tikzpicture}}
\caption{Parametrization of the loop momenta for the 1SF integral topologies $\diagramnumberingsign4$ and $\diagramnumberingsign5$ of table~\ref{tab: 4-loop results 0SF and 1SF}, which depend on the three-loop K3 surface in the odd-parity sector.}
\label{fig: diag_K3_4_and_5}
\end{figure}

We begin with the 1SF integral topologies $\diagramnumberingsign4$ and $\diagramnumberingsign5$ of table~\ref{tab: 4-loop results 0SF and 1SF} in the odd-parity sector. Applying the parametrization in fig.~\ref{fig: diag_K3_4_and_5} for each integral topology, we are able to choose identical sets of four ISPs $z_{12}=k_2^2$, $z_{13}=k_4^2$, $z_{14}=2u_1\cdot k_3$ and $z_{15}=2u_2\cdot k_4$, finding that 
\begin{align}
& \LS \left( \begin{tikzpicture}[baseline={([yshift=-0.1cm]current bounding box.center)}] 
	\node[] (a) at (0,0) {};
	\node[] (a1) at (0.5,0) {};
	\node[] (a2) at (1,0) {};
    \node[] (a3) at (1.5,0) {};
	\node[] (b) at (0,-0.5) {};
	\node[] (b1) at (0.75,-0.5) {};
	\node[] (b2) at (0.75,-0.5) {};
    \node[] (b3) at (1.5,-0.5) {};
	\node[] (c) at (0,-1) {};
	\node[] (c1) at (0.5,-1) {};
	\node[] (c2) at (1,-1) {};
    \node[] (c3) at (1.5,-1) {};
	\node[] (p1) at ($(a)+(-0.2,0)$) {};
	\node[] (p2) at ($(c)+(-0.2,0)$) {};
	\node[] (p3) at ($(c3)+(0.2,0)$) {};
	\node[] (p4) at ($(a3)+(0.2,0)$) {};
    \draw[line width=0.5mm] (p1.center) -- (p4.center);
	\draw[line width=0.5mm] (p2.center) -- (p3.center);
    \draw[line width=0.15mm] (c.center) -- (a.center);
	\draw[line width=0.15mm] (b.center) -- (b1.center);
	\draw[line width=0.15mm] (b1.center) -- (b2.center);
    \draw[line width=0.15mm] (b1.center) -- (a1.center);
    \draw[line width=0.15mm] (b2.center) -- (c3.center);
	\draw[line width=0.15mm] (b2.center) -- (a2.center);
    \draw[line width=0.15mm] (a3.center) -- (c3.center);
    \draw[line width=0.5mm] (a2.center) -- (a3.center);
\end{tikzpicture} {\times} \, 2u_1 {\cdot} k_3  \right) =  
\LS \left( \begin{tikzpicture}[baseline={([yshift=-0.1cm]current bounding box.center)}] 
	\node[] (a) at (0,0) {};
	\node[] (a1) at (0.5,0) {};
	\node[] (a2) at (1,0) {};
    \node[] (a3) at (1.5,0) {};
	\node[] (b) at (0,-0.5) {};
	\node[] (b1) at (0.5,-0.5) {};
	\node[] (b2) at (1,-0.5) {};
    \node[] (b3) at (1.5,-0.5) {};
	\node[] (c) at (0,-1) {};
	\node[] (c1) at (0.5,-1) {};
	\node[] (c2) at (1,-1) {};
    \node[] (c3) at (1.5,-1) {};
	\node[] (p1) at ($(a)+(-0.2,0)$) {};
	\node[] (p2) at ($(c)+(-0.2,0)$) {};
	\node[] (p3) at ($(c3)+(0.2,0)$) {};
	\node[] (p4) at ($(a3)+(0.2,0)$) {};
    \draw[line width=0.5mm] (p1.center) -- (p4.center);
	\draw[line width=0.5mm] (p2.center) -- (p3.center);
    \draw[line width=0.15mm] (c.center) -- (a.center);
	\draw[line width=0.15mm] (b.center) -- (b1.center);
	\draw[line width=0.15mm] (b1.center) -- (b2.center);
    \draw[line width=0.15mm] (b.center) -- (a1.center);
    \draw[line width=0.15mm] (b2.center) -- (c3.center);
	\draw[line width=0.15mm] (b2.center) -- (a2.center);
    \draw[line width=0.15mm] (a3.center) -- (c3.center);
        \draw[line width=0.15mm] (b.center) -- (c.center);
\end{tikzpicture} {\times} \, 2u_1 {\cdot} k_3 \right) \nonumber \\[0.1cm] 
& \, \qquad \qquad \propto \frac{1}{|q|} \, \LS \, \biggl( \int \frac{ x^2 \, d z_{12} \, \cdots \, d z_{15}}{(z_{13}-z_{12})\sqrt{z_{12}} \sqrt{(x^2-1)^2 (z_{13}-1)^2-4x^2z_{15}^2} }
\nonumber \\[0.1cm]
   &  \, \ \ \, \, \qquad \qquad  \times  
\frac{1}{\sqrt{(x^2-1)^2z_{13}^2+x^2z_{14}^2z_{15}^2-2xz_{13}z_{14}(2xz_{14}+(x^2+1)z_{15})}}  \biggr)\,,
\end{align}
where the $z_{14}$ that arises from the ISP in the numerator has canceled against a $z_{14}$ in the denominator. At this stage, we can immediately compute the residue in the integration variable $z_{12}$, followed by the change of variables in eq.~\eqref{eq: change_of_variables_(z-r1)(z-r2)} with respect to $z_{14}$ to rationalize the third square root. This reveals a simple pole in the new integration variable  $t_{14}$, allowing for an additional residue. Thus, we obtain
\begin{align}
& \, \LS \left( \begin{tikzpicture}[baseline={([yshift=-0.1cm]current bounding box.center)}] 
	\node[] (a) at (0,0) {};
	\node[] (a1) at (0.5,0) {};
	\node[] (a2) at (1,0) {};
    \node[] (a3) at (1.5,0) {};
	\node[] (b) at (0,-0.5) {};
	\node[] (b1) at (0.75,-0.5) {};
	\node[] (b2) at (0.75,-0.5) {};
    \node[] (b3) at (1.5,-0.5) {};
	\node[] (c) at (0,-1) {};
	\node[] (c1) at (0.5,-1) {};
	\node[] (c2) at (1,-1) {};
    \node[] (c3) at (1.5,-1) {};
	\node[] (p1) at ($(a)+(-0.2,0)$) {};
	\node[] (p2) at ($(c)+(-0.2,0)$) {};
	\node[] (p3) at ($(c3)+(0.2,0)$) {};
	\node[] (p4) at ($(a3)+(0.2,0)$) {};
    \draw[line width=0.5mm] (p1.center) -- (p4.center);
	\draw[line width=0.5mm] (p2.center) -- (p3.center);
    \draw[line width=0.15mm] (c.center) -- (a.center);
	\draw[line width=0.15mm] (b.center) -- (b1.center);
	\draw[line width=0.15mm] (b1.center) -- (b2.center);
    \draw[line width=0.15mm] (b1.center) -- (a1.center);
    \draw[line width=0.15mm] (b2.center) -- (c3.center);
	\draw[line width=0.15mm] (b2.center) -- (a2.center);
    \draw[line width=0.15mm] (a3.center) -- (c3.center);
    \draw[line width=0.5mm] (a2.center) -- (a3.center);
\end{tikzpicture} {\times} \, 2u_1 {\cdot} k_3  \right) = \, \LS \left( \begin{tikzpicture}[baseline={([yshift=-0.1cm]current bounding box.center)}] 
	\node[] (a) at (0,0) {};
	\node[] (a1) at (0.5,0) {};
	\node[] (a2) at (1,0) {};
    \node[] (a3) at (1.5,0) {};
	\node[] (b) at (0,-0.5) {};
	\node[] (b1) at (0.5,-0.5) {};
	\node[] (b2) at (1,-0.5) {};
    \node[] (b3) at (1.5,-0.5) {};
	\node[] (c) at (0,-1) {};
	\node[] (c1) at (0.5,-1) {};
	\node[] (c2) at (1,-1) {};
    \node[] (c3) at (1.5,-1) {};
	\node[] (p1) at ($(a)+(-0.2,0)$) {};
	\node[] (p2) at ($(c)+(-0.2,0)$) {};
	\node[] (p3) at ($(c3)+(0.2,0)$) {};
	\node[] (p4) at ($(a3)+(0.2,0)$) {};
    \draw[line width=0.5mm] (p1.center) -- (p4.center);
	\draw[line width=0.5mm] (p2.center) -- (p3.center);
    \draw[line width=0.15mm] (c.center) -- (a.center);
	\draw[line width=0.15mm] (b.center) -- (b1.center);
	\draw[line width=0.15mm] (b1.center) -- (b2.center);
    \draw[line width=0.15mm] (b.center) -- (a1.center);
    \draw[line width=0.15mm] (b2.center) -- (c3.center);
	\draw[line width=0.15mm] (b2.center) -- (a2.center);
    \draw[line width=0.15mm] (a3.center) -- (c3.center);
        \draw[line width=0.15mm] (b.center) -- (c.center);
\end{tikzpicture} {\times} \, 2u_1 {\cdot} k_3 \right) \nonumber \\[0.1cm] 
& \qquad \quad \propto \ \frac{x}{|q|} \, \LS \biggl( \int \frac{d z_{13}d z_{15}}{\sqrt{z_{13}} \sqrt{4 z_{13} - z_{15}^2} \sqrt{(x^2-1)^2 (z_{13}-1)^2 - 4 x^2 z_{15}^2}} \biggr).
\label{eq: diag_4 + diag_5 LS}
\end{align}
At this point, trying to rationalize further the square roots does not reveal any simple poles at which to take residues. Instead, we rationalize the first square root by introducing $z_{13}=t^2_{13}$. Then, we rationalize the second square root by introducing the change of variables in eq.~\eqref{eq: change_of_variables_(z-r1)(z-r2)} from $z_{15}$ to $t_{15}$, leaving us with a single square root in the denominator. Relabeling $\{ t_{13}, t_{15}\} \to \{ t_1, t_2 \}$, the leading singularity becomes
\begin{align}
\nonumber
&\LS \left( \begin{tikzpicture}[baseline={([yshift=-0.1cm]current bounding box.center)}] 
	\node[] (a) at (0,0) {};
	\node[] (a1) at (0.5,0) {};
	\node[] (a2) at (1,0) {};
    \node[] (a3) at (1.5,0) {};
	\node[] (b) at (0,-0.5) {};
	\node[] (b1) at (0.75,-0.5) {};
	\node[] (b2) at (0.75,-0.5) {};
    \node[] (b3) at (1.5,-0.5) {};
	\node[] (c) at (0,-1) {};
	\node[] (c1) at (0.5,-1) {};
	\node[] (c2) at (1,-1) {};
    \node[] (c3) at (1.5,-1) {};
	\node[] (p1) at ($(a)+(-0.2,0)$) {};
	\node[] (p2) at ($(c)+(-0.2,0)$) {};
	\node[] (p3) at ($(c3)+(0.2,0)$) {};
	\node[] (p4) at ($(a3)+(0.2,0)$) {};
    \draw[line width=0.5mm] (p1.center) -- (p4.center);
	\draw[line width=0.5mm] (p2.center) -- (p3.center);
    \draw[line width=0.15mm] (c.center) -- (a.center);
	\draw[line width=0.15mm] (b.center) -- (b1.center);
	\draw[line width=0.15mm] (b1.center) -- (b2.center);
    \draw[line width=0.15mm] (b1.center) -- (a1.center);
    \draw[line width=0.15mm] (b2.center) -- (c3.center);
	\draw[line width=0.15mm] (b2.center) -- (a2.center);
    \draw[line width=0.15mm] (a3.center) -- (c3.center);
    \draw[line width=0.5mm] (a2.center) -- (a3.center);
\end{tikzpicture} {\times} \, 2u_1 {\cdot} k_3 \right) \; =\; 
\LS \left( \begin{tikzpicture}[baseline={([yshift=-0.1cm]current bounding box.center)}] 
	\node[] (a) at (0,0) {};
	\node[] (a1) at (0.5,0) {};
	\node[] (a2) at (1,0) {};
    \node[] (a3) at (1.5,0) {};
	\node[] (b) at (0,-0.5) {};
	\node[] (b1) at (0.5,-0.5) {};
	\node[] (b2) at (1,-0.5) {};
    \node[] (b3) at (1.5,-0.5) {};
	\node[] (c) at (0,-1) {};
	\node[] (c1) at (0.5,-1) {};
	\node[] (c2) at (1,-1) {};
    \node[] (c3) at (1.5,-1) {};
	\node[] (p1) at ($(a)+(-0.2,0)$) {};
	\node[] (p2) at ($(c)+(-0.2,0)$) {};
	\node[] (p3) at ($(c3)+(0.2,0)$) {};
	\node[] (p4) at ($(a3)+(0.2,0)$) {};
    \draw[line width=0.5mm] (p1.center) -- (p4.center);
	\draw[line width=0.5mm] (p2.center) -- (p3.center);
    \draw[line width=0.15mm] (c.center) -- (a.center);
	\draw[line width=0.15mm] (b.center) -- (b1.center);
	\draw[line width=0.15mm] (b1.center) -- (b2.center);
    \draw[line width=0.15mm] (b.center) -- (a1.center);
    \draw[line width=0.15mm] (b2.center) -- (c3.center);
	\draw[line width=0.15mm] (b2.center) -- (a2.center);
    \draw[line width=0.15mm] (a3.center) -- (c3.center);
        \draw[line width=0.15mm] (b.center) -- (c.center);
\end{tikzpicture} {\times} \, 2u_1 {\cdot} k_3 \right)\\ 
\ \ & \propto \frac{x}{|q|} \, \int\! \frac{d t_1 d t_2}{\sqrt{t_2^2 (t_1^2-1)^2 (x^2-1)^2 - 4 x^2 t_1^2 (t_2^2+1)^2}} \equiv \frac{x}{|q|} \, \int\! \frac{ dt_1 dt_2}{\sqrt{P_6(t_1,t_2)}}\,.
\label{eq: 3-loop K3}
\end{align}
The polynomial $P_6(t_1,t_2)$ has total degree six, and degree four in each variable. Following the discussion in sec.~\ref{sec:review_DE_LS} below eq.~\eqref{eq: LS_definition}, this defines an integral over a K3 surface. Notably, in this representation we find exactly the same degree-6 polynomial as in the three-loop analysis of ref.~\cite{Frellesvig:2024zph}, thus confirming that the geometry is the same. Similarly, we also verified that the integrals are annihilated by the same irreducible third-order Picard--Fuchs operator; see refs.~\cite{Ruf:2021egk,Dlapa:2022wdu,Frellesvig:2024zph}.

\begin{figure}[t]
\centering
\begin{tikzpicture}[baseline=(current bounding box.center)] 
	\node[] (a) at (0,0) {};
	\node[] (a1) at (2,0) {};
	\node[] (a2) at (4,0) {};
	\node[] (a3) at (6,0) {};
	\node[] (b) at (0,-2) {};
	\node[] (b1) at (6,-2) {};
	\node[] (c) at (0,-4) {};
	\node[] (c1) at (6,-4) {};
	\node[] (p1) at ($(a)+(-0.5,0)$) {};
	\node[] (p2) at ($(c)+(-0.5,0)$) {};
	\node[] (p3) at ($(c1)+(0.5,0)$) {};
	\node[] (p4) at ($(a3)+(0.5,0)$) {};
	\draw[line width=0.15mm, postaction={decorate}] (b.center) -- node[sloped, allow upside down, label={[xshift=0.15cm, yshift=0cm]$k_1$}] {\midarrow} (a.center);
	\draw[line width=0.15mm, postaction={decorate}] (a1.center) -- node[sloped, allow upside down, label={[xshift=-0.05cm, yshift=0.45cm]$k_1{-}k_2$}] {\midarrow} (b.center);
	\draw[line width=0.15mm, postaction={decorate}] (a2.center) -- node[sloped, allow upside down, label={[xshift=-1.6cm, yshift=-0.45cm]$k_2{-}k_3$}] {\midarrow} (b1.center);
	\draw[line width=0.15mm, postaction={decorate}] (a3.center) -- node[sloped, allow upside down, label={[xshift=-0.15cm, yshift=0cm]$k_3{-}q$}] {\midarrow} (b1.center);
	\draw[line width=0.15mm, postaction={decorate}] (b1.center) -- node[sloped, allow upside down, label={[xshift=0cm, yshift=0.9cm]$k_2{+}k_4$}] {\midarrow} (b.center);
	\draw[line width=0.15mm, postaction={decorate}] (b.center) -- node[sloped, allow upside down, label={[xshift=-0.95cm, yshift=0.05cm]$k_4$}] {\midarrow} (c.center);
	\draw[line width=0.15mm, postaction={decorate}] (c1.center) -- node[sloped, allow upside down, label={[xshift=1.45cm, yshift=0cm]$k_4{+}q$}] {\midarrow} (b1.center);
	\draw[line width=0.5mm, postaction={decorate}] (a.center) -- node[sloped, allow upside down, label={[xshift=0cm, yshift=-0.15cm]$2u_1 {\cdot} k_1$}] {\midarrow} (a1.center);
	\draw[line width=0.5mm, postaction={decorate}] (a1.center) -- node[sloped, allow upside down, label={[xshift=0cm, yshift=-0.15cm]$2u_1 {\cdot} k_2$}] {\midarrow} (a2.center);
	\draw[line width=0.5mm, postaction={decorate}] (a2.center) -- node[sloped, allow upside down, label={[xshift=0cm, yshift=-0.15cm]$2u_1 {\cdot} k_3$}] {\midarrow} (a3.center);
	\draw[line width=0.5mm, postaction={decorate}] (c.center) -- node[sloped, allow upside down, label={[xshift=0cm, yshift=-0.9cm]$2u_2 {\cdot} k_4$}] {\midarrow} (c1.center);
	\draw[line width=0.5mm] (p1.center) -- (p4.center);
	\draw[line width=0.5mm] (p2.center) -- (p3.center);
\end{tikzpicture}
\caption{Parametrization of the loop momenta for the 1SF integral topology $\diagramnumberingsign6$ of table~\ref{tab: 4-loop results 0SF and 1SF}, which depends on the three-loop K3 surface in the even-parity sector.}
\label{fig: diag_K3_6}
\end{figure}

We now turn our attention to the 1SF integral topology $\diagramnumberingsign6$ in table~\ref{tab: 4-loop results 0SF and 1SF}, which we parametrize as in fig.~\ref{fig: diag_K3_6}. In this case, we focus on the even-parity sector, for which we use the integration order $\{ k_1,k_3,k_2,k_4 \}$ and introduce three ISPs, defined as $z_{12}=k_2^2$, $z_{13}=(k_2-q)^2$ and $z_{14}=2u_1 \cdot k_4$. Then, we find the leading singularity of the scalar integral to be 
\begin{align}
\label{eq: Diagram 6 LS}
    \LS 
    \left( 
   \begin{tikzpicture}[baseline={([yshift=-0.1cm]current bounding box.center)}] 
	\node[] (a) at (0,0) {};
	\node[] (a1) at (0.5,0) {};
	\node[] (a2) at (1,0) {};
    \node[] (a3) at (1.5,0) {};
	\node[] (b) at (0,-0.5) {};
	\node[] (b1) at (0.5,-0.5) {};
	\node[] (b2) at (1,-0.5) {};
    \node[] (b3) at (1.5,-0.5) {};
	\node[] (c) at (0,-1) {};
	\node[] (c1) at (0.5,-1) {};
	\node[] (c2) at (1,-1) {};
    \node[] (c3) at (1.5,-1) {};
	\node[] (p1) at ($(a)+(-0.2,0)$) {};
	\node[] (p2) at ($(c)+(-0.2,0)$) {};
	\node[] (p3) at ($(c3)+(0.2,0)$) {};
	\node[] (p4) at ($(a3)+(0.2,0)$) {};
    \draw[line width=0.5mm] (p1.center) -- (p4.center);
	\draw[line width=0.5mm] (p2.center) -- (p3.center);
    \draw[line width=0.15mm] (c.center) -- (a.center);
	\draw[line width=0.15mm] (b.center) -- (b1.center);
	\draw[line width=0.15mm] (b1.center) -- (b2.center);
    \draw[line width=0.15mm] (b.center) -- (a1.center);
    \draw[line width=0.15mm] (b2.center) -- (b3.center);
	\draw[line width=0.15mm] (b3.center) -- (a2.center);
    \draw[line width=0.15mm] (a3.center) -- (c3.center);
\end{tikzpicture} 
    \right)
\propto & \ \frac{1}{q^2} \, \LS \, \biggl( \int  \frac{x\,  d z_{12}d z_{13} d z_{14}}
{\sqrt{z_{12}}\sqrt{z_{13}}\sqrt{(x^2-1)^2-4x^2z_{14}^2}} \nonumber \\
   & \,  \times
 \frac{1}{\sqrt{(z_{12}^2 - 2 z_{12} (z_{13}+1) + (z_{13}-1)^2) z_{14}^2 - 4 z_{12} z_{13}}} \biggr)\,.
\end{align}
By first shifting $z_{12}\to z_{12} +z_{13}$ and subsequently applying the change of variables in eq.~\eqref{eq: change_of_variables_(z-r1)(z-r2)} with respect to the variable $z_{13}$, we can simultaneously rationalize the first two square roots. Then, the integration variable $z_{12}$ appears only quadratically in the last square root. Thus, we can apply again the change of variables from eq.~\eqref{eq: change_of_variables_(z-r1)(z-r2)} to rationalize it, exposing a simple pole at $t_{12}=0$. Taking the residue at this pole, we obtain
\begin{equation}
    \LS 
    \left( 
   \begin{tikzpicture}[baseline={([yshift=-0.1cm]current bounding box.center)}] 
	\node[] (a) at (0,0) {};
	\node[] (a1) at (0.5,0) {};
	\node[] (a2) at (1,0) {};
    \node[] (a3) at (1.5,0) {};
	\node[] (b) at (0,-0.5) {};
	\node[] (b1) at (0.5,-0.5) {};
	\node[] (b2) at (1,-0.5) {};
    \node[] (b3) at (1.5,-0.5) {};
	\node[] (c) at (0,-1) {};
	\node[] (c1) at (0.5,-1) {};
	\node[] (c2) at (1,-1) {};
    \node[] (c3) at (1.5,-1) {};
	\node[] (p1) at ($(a)+(-0.2,0)$) {};
	\node[] (p2) at ($(c)+(-0.2,0)$) {};
	\node[] (p3) at ($(c3)+(0.2,0)$) {};
	\node[] (p4) at ($(a3)+(0.2,0)$) {};
    \draw[line width=0.5mm] (p1.center) -- (p4.center);
	\draw[line width=0.5mm] (p2.center) -- (p3.center);
    \draw[line width=0.15mm] (c.center) -- (a.center);
	\draw[line width=0.15mm] (b.center) -- (b1.center);
	\draw[line width=0.15mm] (b1.center) -- (b2.center);
    \draw[line width=0.15mm] (b.center) -- (a1.center);
    \draw[line width=0.15mm] (b2.center) -- (b3.center);
	\draw[line width=0.15mm] (b3.center) -- (a2.center);
    \draw[line width=0.15mm] (a3.center) -- (c3.center);
\end{tikzpicture} 
    \right)
\propto \frac{x}{q^2} \, \LS \biggl( \int  \frac{d t_{13} d z_{14}}
{\sqrt{(x^2-1)^2-4x^2z_{14}^2} \sqrt{t_{13}^2 (4 z_{14}^2+2)- t_{13}^4-1}} \biggr).
\end{equation}
Now, we can rationalize the first square root via the change of variables in eq.~\eqref{eq: change_of_variables_(z-r1)(z-r2)} with respect to $z_{14}$. After relabeling $t_{13} \to i\, t_2$ and $t_{14} \to i\, t_1$, we arrive at the expression
\begin{equation}
    \LS 
    \left( 
   \begin{tikzpicture}[baseline={([yshift=-0.1cm]current bounding box.center)}] 
	\node[] (a) at (0,0) {};
	\node[] (a1) at (0.5,0) {};
	\node[] (a2) at (1,0) {};
    \node[] (a3) at (1.5,0) {};
	\node[] (b) at (0,-0.5) {};
	\node[] (b1) at (0.5,-0.5) {};
	\node[] (b2) at (1,-0.5) {};
    \node[] (b3) at (1.5,-0.5) {};
	\node[] (c) at (0,-1) {};
	\node[] (c1) at (0.5,-1) {};
	\node[] (c2) at (1,-1) {};
    \node[] (c3) at (1.5,-1) {};
	\node[] (p1) at ($(a)+(-0.2,0)$) {};
	\node[] (p2) at ($(c)+(-0.2,0)$) {};
	\node[] (p3) at ($(c3)+(0.2,0)$) {};
	\node[] (p4) at ($(a3)+(0.2,0)$) {};
    \draw[line width=0.5mm] (p1.center) -- (p4.center);
	\draw[line width=0.5mm] (p2.center) -- (p3.center);
    \draw[line width=0.15mm] (c.center) -- (a.center);
	\draw[line width=0.15mm] (b.center) -- (b1.center);
	\draw[line width=0.15mm] (b1.center) -- (b2.center);
    \draw[line width=0.15mm] (b.center) -- (a1.center);
    \draw[line width=0.15mm] (b2.center) -- (b3.center);
	\draw[line width=0.15mm] (b3.center) -- (a2.center);
    \draw[line width=0.15mm] (a3.center) -- (c3.center);
\end{tikzpicture} 
    \right)
\propto \frac{x}{q^2} \, \int\! \frac{d t_1 d t_2}{\sqrt{t_2^2 (t_1^2-1)^2 (x^2-1)^2 - 4 x^2 t_1^2 (t_2^2+1)^2}}\,.
\end{equation}
Comparing with eq.~\eqref{eq: 3-loop K3}, we find that the even-parity sector for integral topology $\diagramnumberingsign6$ also depends on the same K3 surface, since the leading singularity yields a period integral over the same degree-6 polynomial.

\begin{figure}[t]
\centering
\subfloat[]{\begin{tikzpicture}[baseline=(current bounding box.center), scale=0.82] 
        \node[] (a) at (0,0) {};
        \node[] (a1) at (3,0) {};
        \node[] (a2) at (6,0) {};
        \node[] (b) at (0,-2) {};
        \node[] (c) at (0,-4) {};
        \node[] (c1) at (3,-4) {};
        \node[] (c2) at (6,-4) {};
        \node[] (p1) at ($(a)+(-0.5,0)$) {};
        \node[] (p2) at ($(c)+(-0.5,0)$) {};
        \node[] (p3) at ($(c2)+(0.5,0)$) {};
        \node[] (p4) at ($(a2)+(0.5,0)$) {};
        \draw[line width=0.15mm, postaction={decorate}] (b.center) -- node[sloped, allow upside down, label={[xshift=0.2cm, yshift=0cm]$k_1$}] {\midarrow} (a.center);
        \draw[line width=0.15mm, postaction={decorate}] (a1.center) -- node[sloped, allow upside down, label={[xshift=0cm, yshift=0.4cm]$k_1{-}k_4$}] {\midarrow} (b.center);
        \draw[line width=0.15mm, postaction={decorate}] (a2.center) -- node[sloped, allow upside down, label={[xshift=-0.2cm, yshift=0cm]$k_4{-}q$}] {\midarrow} (c2.center);
        \draw[line width=0.15mm, postaction={decorate}] (c2.center) -- node[sloped, allow upside down, label={[xshift=1.44cm, yshift=0.75cm]$k_3{+}k_4$}] {\midarrow} (b.center);
        \draw[line width=0.15mm, postaction={decorate}] (c1.center) -- node[sloped, allow upside down, label={[xshift=0.4cm, yshift=0.2cm]$k_2{-}k_3$}] {\midarrow} (b.center);
        \draw[line width=0.15mm, postaction={decorate}] (b.center) -- node[sloped, allow upside down, label={[xshift=-0.95cm, yshift=0.05cm]$k_2$}] {\midarrow} (c.center);
        \draw[line width=0.5mm, postaction={decorate}] (a.center) -- node[sloped, allow upside down, label={[xshift=0cm, yshift=-0.15cm]$2u_1 {\cdot} k_1$}] {\midarrow} (a1.center);
        \draw[line width=0.5mm, postaction={decorate}] (a1.center) -- node[sloped, allow upside down, label={[xshift=0cm, yshift=-0.15cm]$2u_1 {\cdot} k_4$}] {\midarrow} (a2.center);
        \draw[line width=0.5mm, postaction={decorate}] (c.center) -- node[sloped, allow upside down, label={[xshift=0cm, yshift=-0.9cm]$2u_2 {\cdot} k_2$}] {\midarrow} (c1.center);
        \draw[line width=0.5mm, postaction={decorate}] (c1.center) -- node[sloped, allow upside down, label={[xshift=0cm, yshift=-0.9cm]$2u_2 {\cdot} k_3$}] {\midarrow} (c2.center);
        \draw[line width=0.15mm, white] (3,0) arc (0:180:1.5);
        \node[label={[xshift=0cm, yshift=-0.05cm]$\textcolor{white}{k_1}$}] (kwhite) at (1.5,1.5) {};
        \draw[line width=0.5mm] (p1.center) -- node[] {} (a.center);
        \draw[line width=0.5mm] (a2.center) -- node[] {} (p4.center);
        \draw[line width=0.5mm] (p2.center) -- node[] {} (c.center);
        \draw[line width=0.5mm] (c2.center) -- node[] {} (p3.center);
    \end{tikzpicture}} \quad \enspace \subfloat[]{\begin{tikzpicture}[baseline=(current bounding box.center), scale=0.82] 
        \node[] (a) at (0,0) {};
        \node[] (a1) at (3,0) {};
        \node[] (a2) at (6,0) {};
        \node[] (b) at (0,-4) {};
        \node[] (b1) at (3,-4) {};
        \node[] (b2) at (6,-4) {};
        \node[] (x) at (4.5,-2) {};
        \node[] (p1) at ($(a)+(-0.5,0)$) {};
        \node[] (p2) at ($(b)+(-0.5,0)$) {};
        \node[] (p3) at ($(b2)+(0.5,0)$) {};
        \node[] (p4) at ($(a2)+(0.5,0)$) {};
        \draw[line width=0.5mm, postaction={decorate}] (a.center) -- node[sloped, allow upside down, label={[xshift=0.1cm, yshift=-0.95cm]$2u_1{\cdot}(k_1{-}k_4)$}] {\midarrow} (a1.center);
        \draw[line width=0.15mm, postaction={decorate}] (a.center) -- node[sloped, allow upside down, label={[xshift=-1.475cm, yshift=0cm]$k_4{-}q$}] {\midarrow} (b.center);
        \draw[line width=0.15mm, postaction={decorate}] (x.center) -- node[sloped, allow upside down, label={[xshift=0.25cm, yshift=0.2cm]$k_2{+}k_4$}] {\midarrow} (a1.center);
        \draw[line width=0.15mm, postaction={decorate}] (a2.center) -- node[sloped, allow upside down, label={[xshift=-0.15cm, yshift=0.25cm]$k_2$}] {\midarrow} (x.center);
        \draw[line width=0.15mm, postaction={decorate}] (x.center) -- node[sloped, allow upside down, label={[xshift=-1.55cm, yshift=0.55cm]$k_3{-}k_4$}] {\midarrow} (b1.center);
        \draw[line width=0.15mm, postaction={decorate}] (b2.center) -- node[sloped, allow upside down, label={[xshift=0.9cm, yshift=0.5cm]$k_3$}] {\midarrow} (x.center);
        \draw[line width=0.5mm, postaction={decorate}] (b.center) -- node[sloped, allow upside down, label={[xshift=0cm, yshift=-0.9cm]$2u_2 {\cdot} k_4$}] {\midarrow} (b1.center);
        \draw[line width=0.5mm, postaction={decorate}] (b1.center) -- node[sloped, allow upside down, label={[xshift=0cm, yshift=-0.9cm]$2u_2 {\cdot} k_3$}] {\midarrow} (b2.center);
        \draw[line width=0.5mm, postaction={decorate}] (a1.center) -- node[sloped, allow upside down, label={[xshift=0cm, yshift=-0.15cm]$2u_1 {\cdot} k_2$}] {\midarrow} (a2.center);
        \draw[line width=0.15mm] (3,0) arc (0:180:1.5);
        \draw[line width=0.15mm, postaction={decorate}] (1.5,1.5) -- node[sloped, allow upside down, label={[xshift=0cm, yshift=0.9cm]$k_1$}] {\midarrow} (1.4995,1.5);
        \draw[line width=0.5mm] (p1.center) -- node[] {} (a.center);
        \draw[line width=0.5mm] (a2.center) -- node[] {} (p4.center);
        \draw[line width=0.5mm] (p2.center) -- node[] {} (b.center);
        \draw[line width=0.5mm] (b2.center) -- node[] {} (p3.center);
    \end{tikzpicture}}
\caption{Parametrization of the loop momenta for the 2SF integral topologies $\diagramnumberingsign38$ and $\diagramnumberingsign39$ of table~\ref{tab: 4-loop results 2SF}, which depend on the three-loop K3 surface in the odd-parity sector.}
\label{fig: diag_K3_38_and_39}
\end{figure}

Next, let us consider the odd-parity sector for the 2SF integral topology $\diagramnumberingsign38$ in table~\ref{tab: 4-loop results 2SF}, with the parametrization shown in fig.~\ref{fig: diag_K3_38_and_39}(a). In this case we require three ISPs, given by $z_{11}=k_3^2$, $z_{12}=k_4^2$ and $z_{13}=2 u_2 \cdot k_4$. Adding a dot to one of the parity-odd matter propagators, we obtain the leading singularity
{\allowdisplaybreaks
\begin{align}
    & \, \LS
    \left( 
    \begin{tikzpicture}[baseline={([yshift=-0.1cm]current bounding box.center)}] 
	\node[] (a) at (0,0) {};
	\node[] (a1) at (0.5,0) {};
	\node[] (a2) at (1,0) {};
    \node[] (a3) at (1.5,0) {};
	\node[] (b) at (0,-0.5) {};
	\node[] (b1) at (0.5,-0.5) {};
	\node[] (b2) at (1,-0.5) {};
    \node[] (b3) at (1.5,-0.5) {};
	\node[] (c) at (0,-1) {};
	\node[] (c1) at (0.5,-1) {};
	\node[] (c2) at (0.75,-1) {};
    \node[] (c3) at (1.5,-1) {};
	\node[] (p1) at ($(a)+(-0.2,0)$) {};
	\node[] (p2) at ($(c)+(-0.2,0)$) {};
	\node[] (p3) at ($(c3)+(0.2,0)$) {};
	\node[] (p4) at ($(a3)+(0.2,0)$) {};
    \draw[line width=0.5mm] (p1.center) -- (p4.center);
	\draw[line width=0.5mm] (p2.center) -- (p3.center);
    \draw[line width=0.15mm] (c.center) -- (a.center);
	\draw[line width=0.15mm] (b.center) -- (a2.center);
    \draw[line width=0.15mm] (b.center) -- (c3.center);
    \draw[line width=0.15mm] (c2.center) -- (b.center);
    \draw[line width=0.15mm] (a3.center) -- (c3.center);
    \node at (1,-1) [circle,fill,inner sep=1.5pt]{};
\end{tikzpicture}
    \right) \propto \nonumber \\
& \propto \frac{1}{|q|} \, \LS \! \left(\int \! \frac{\varepsilon\,x\, (z_{11}+z_{12})\, z_{13} \, dz_{11} dz_{12}dz_{13}}{\sqrt{z_{11}}\sqrt{z_{12}}\sqrt{4z_{12}{-}z_{13}^2}\sqrt{(x^2{-}1)^2(z_{12}{-}1)^2{-}4x^2z_{13}^2}\left((z_{11}{-}z_{12})^2{+}z_{11}z_{13}^2\right)} \right) \! .
\end{align}
}
Changing variables to $z_{11}=t_{11}^2$ and taking the residue in $t_{11}$ yields 
\begin{align}
\label{eq: Diagram 38 LS}
    \LS
    \left( 
    \begin{tikzpicture}[baseline={([yshift=-0.1cm]current bounding box.center)}] 
	\node[] (a) at (0,0) {};
	\node[] (a1) at (0.5,0) {};
	\node[] (a2) at (1,0) {};
    \node[] (a3) at (1.5,0) {};
	\node[] (b) at (0,-0.5) {};
	\node[] (b1) at (0.5,-0.5) {};
	\node[] (b2) at (1,-0.5) {};
    \node[] (b3) at (1.5,-0.5) {};
	\node[] (c) at (0,-1) {};
	\node[] (c1) at (0.5,-1) {};
	\node[] (c2) at (0.75,-1) {};
    \node[] (c3) at (1.5,-1) {};
	\node[] (p1) at ($(a)+(-0.2,0)$) {};
	\node[] (p2) at ($(c)+(-0.2,0)$) {};
	\node[] (p3) at ($(c3)+(0.2,0)$) {};
	\node[] (p4) at ($(a3)+(0.2,0)$) {};
    \draw[line width=0.5mm] (p1.center) -- (p4.center);
	\draw[line width=0.5mm] (p2.center) -- (p3.center);
    \draw[line width=0.15mm] (c.center) -- (a.center);
	\draw[line width=0.15mm] (b.center) -- (a2.center);
    \draw[line width=0.15mm] (b.center) -- (c3.center);
    \draw[line width=0.15mm] (c2.center) -- (b.center);
    \draw[line width=0.15mm] (a3.center) -- (c3.center);
    \node at (1,-1) [circle,fill,inner sep=1.5pt]{};
\end{tikzpicture}
    \right)
\propto & \ \frac{1}{|q|} \, \LS \left( \rule{0cm}{0.9cm} \right. \int \frac{\varepsilon\,x\, dz_{12}dz_{13}}
{(4z_{12}-z_{13}^2)\sqrt{z_{12}}\sqrt{(x^2-1)^2(z_{12}-1)^2-4x^2z_{13}^2}} \nonumber \\
& \ \times \, \frac{4z_{12}-z_{13}^2 + z_{13}\sqrt{z_{13}^2-4z_{12}}}{\sqrt{2z_{12}-z_{13}\big(z_{13} - \sqrt{z_{13}^2-4z_{12}}\big)}} \left. \rule{0cm}{0.9cm} \right) \, .
\end{align}
The numerator and denominator in the last fraction can be further simplified, resulting in
\begin{equation}
    \LS
    \left( 
    \begin{tikzpicture}[baseline={([yshift=-0.1cm]current bounding box.center)}] 
	\node[] (a) at (0,0) {};
	\node[] (a1) at (0.5,0) {};
	\node[] (a2) at (1,0) {};
    \node[] (a3) at (1.5,0) {};
	\node[] (b) at (0,-0.5) {};
	\node[] (b1) at (0.5,-0.5) {};
	\node[] (b2) at (1,-0.5) {};
    \node[] (b3) at (1.5,-0.5) {};
	\node[] (c) at (0,-1) {};
	\node[] (c1) at (0.5,-1) {};
	\node[] (c2) at (0.75,-1) {};
    \node[] (c3) at (1.5,-1) {};
	\node[] (p1) at ($(a)+(-0.2,0)$) {};
	\node[] (p2) at ($(c)+(-0.2,0)$) {};
	\node[] (p3) at ($(c3)+(0.2,0)$) {};
	\node[] (p4) at ($(a3)+(0.2,0)$) {};
    \draw[line width=0.5mm] (p1.center) -- (p4.center);
	\draw[line width=0.5mm] (p2.center) -- (p3.center);
    \draw[line width=0.15mm] (c.center) -- (a.center);
	\draw[line width=0.15mm] (b.center) -- (a2.center);
    \draw[line width=0.15mm] (b.center) -- (c3.center);
    \draw[line width=0.15mm] (c2.center) -- (b.center);
    \draw[line width=0.15mm] (a3.center) -- (c3.center);
    \node at (1,-1) [circle,fill,inner sep=1.5pt]{};
\end{tikzpicture}
    \right)
\propto  \frac{\varepsilon \, x}{|q|} \, \LS \left( \int \frac{dz_{12}dz_{13}}
{\sqrt{z_{12}}\sqrt{4z_{12}-z_{13}^2}\sqrt{(x^2-1)^2(z_{12}-1)^2-4x^2z_{13}^2}} \right).
\end{equation}
Comparing with eq.~\eqref{eq: diag_4 + diag_5 LS}, we see that the expressions are equivalent. Therefore, the leading singularity is a period integral over the same K3 surface that we found in eq.~\eqref{eq: 3-loop K3}.

Finally, we turn to the odd-parity sector of the 2SF integral topology $\diagramnumberingsign39$ in table~\ref{tab: 4-loop results 2SF}. Parametrizing it as in fig.~\ref{fig: diag_K3_38_and_39}(b), we need to introduce two ISPs, defined as $z_{11}=k_4^2$ and $z_{12}=2 u_1 \cdot k_4$. In this case, dotting the matter propagator inside of the bubble, we directly obtain
\begin{equation}
        \LS \left(\begin{tikzpicture}[baseline={([yshift=-0.1cm]current bounding box.center)}]
    \node[] (a) at (0,0) {};
    \node[] (a1) at (0.8,0) {};
    \node[] (a2) at (1,0) {};
    \node[] (a3) at (1.5,0) {};
    \node[] (b1) at (1,-0.75) {};
    \node[] (c) at (0,-1) {};
    \node[] (c1) at (0.8,-1) {};
    \node[] (c2) at (1.25,-1) {};
    \node[] (c3) at (1.5,-1) {};
    \node[] (p1) at ($(a)+(-0.2,0)$) {};
	\node[] (p2) at ($(c)+(-0.2,0)$) {};
	\node[] (p3) at ($(c3)+(0.2,0)$) {};
	\node[] (p4) at ($(a3)+(0.2,0)$) {};
    \draw[line width=0.5mm] (p1.center) -- (p4.center);
	\draw[line width=0.5mm] (p2.center) -- (p3.center);
    \draw[line width=0.15mm] (a.center) -- (c.center);
    \draw[line width=0.15mm] (a3.center) -- (c1.center);
    \draw[line width=0.15mm] (a1.center) -- (c3.center);
    \draw[line width=0.15mm] (0.8,0) arc (0:180:0.4);
    \node at (0.4,0) [circle,fill,inner sep=1.5pt]{};
\end{tikzpicture} \right)
\propto \frac{x}{|q|} \, \LS \left( \int \frac{d z_{11} d z_{12} }{\sqrt{z_{11}} \sqrt{4z_{11}-z_{12}^2} \sqrt{(x^2-1)^2(z_{11}-1)^2-4x^2z_{12}^2}} \right).
\end{equation}
Again, this expression is equivalent to eq.~\eqref{eq: diag_4 + diag_5 LS} and thus depends on the same K3 surface, defined in eq.~\eqref{eq: 3-loop K3}. For all of the previous integrals, we verified that the associated third-order Picard--Fuchs operator is equal to the operator annihilating the three-loop integral over the K3 surface~\cite{Ruf:2021egk,Dlapa:2022wdu,Frellesvig:2024zph}, thus confirming that the geometry is the same.

\subsubsection{A second K3 surface at 2SF order}
\label{sec:K32}

In this section, we conclude the classification of the non-trivial geometries which appear at the 5PM order by studying the 2SF integral topology $\diagramnumberingsign40$ in table~\ref{tab: 4-loop results 2SF}. In ref.~\cite{MichaelAmplitudes}, it was already suggested that the even-parity sector for this topology would lie beyond polylogarithms, and it was afterwards investigated via differential equations in ref.~\cite{Klemm:2024wtd}, where an irreducible third-order Picard--Fuchs operator was found. In the following, we find an explicit representation for the leading singularity as a period integral over the K3 surface only via residues and rationalization.%
\footnote{See also ref.~\cite{Duhr:2025lbz} for an alternative approach via the leading singularities method.}

\begin{figure}[t]
\centering
\begin{tikzpicture}[baseline=(current bounding box.center)] 
	\node[] (a) at (0,0) {};
	\node[] (a1) at (3,0) {};
	\node[] (a2) at (6,0) {};
	\node[] (b) at (3,-2) {};
	\node[] (c) at (0,-4) {};
	\node[] (c1) at (3,-4) {};
	\node[] (c2) at (6,-4) {};
	\node[] (p1) at ($(a)+(-0.5,0)$) {};
	\node[] (p2) at ($(c)+(-0.5,0)$) {};
	\node[] (p3) at ($(c2)+(0.5,0)$) {};
	\node[] (p4) at ($(a2)+(0.5,0)$) {};
	\draw[line width=0.15mm, postaction={decorate}] (b.center) -- node[sloped, allow upside down, label={[xshift=0.2cm, yshift=0.25cm]$k_1$}] {\midarrow} (a.center);
	\draw[line width=0.15mm, postaction={decorate}] (a1.center) -- node[sloped, allow upside down, label={[xshift=-0.15cm, yshift=0cm]$k_1{-}k_2$}] {\midarrow} (b.center);
	\draw[line width=0.15mm, postaction={decorate}] (b.center) -- node[sloped, allow upside down, label={[xshift=-0.95cm, yshift=0cm]$k_4$}] {\midarrow} (c1.center);
	\draw[line width=0.15mm, postaction={decorate}] (a2.center) -- node[sloped, allow upside down, label={[xshift=-0.15cm, yshift=0cm]$k_2{-}k_3{-}q$}] {\midarrow} (c2.center);
	\draw[line width=0.15mm, postaction={decorate}] (a.center) -- node[sloped, allow upside down, label={[xshift=-0.95cm, yshift=0cm]$k_3$}] {\midarrow} (c.center);
	\draw[line width=0.15mm, postaction={decorate}] (c2.center) -- node[sloped, allow upside down, label={[xshift=1.3cm, yshift=0.8cm]$k_2{+}k_4$}] {\midarrow} (b.center);
	\draw[line width=0.5mm, postaction={decorate}] (a.center) -- node[sloped, allow upside down, label={[xshift=0cm, yshift=-0.15cm]$2u_1 {\cdot} (k_1{-}k_3)$}] {\midarrow} (a1.center);
	\draw[line width=0.5mm, postaction={decorate}] (a1.center) -- node[sloped, allow upside down, label={[xshift=0cm, yshift=-0.15cm]$2u_1 {\cdot} (k_2{-}k_3)$}] {\midarrow} (a2.center);
	\draw[line width=0.5mm, postaction={decorate}] (c.center) -- node[sloped, allow upside down, label={[xshift=0cm, yshift=-0.9cm]$2u_2 {\cdot} k_3$}] {\midarrow} (c1.center);
	\draw[line width=0.5mm, postaction={decorate}] (c1.center) -- node[sloped, allow upside down, label={[xshift=0cm, yshift=-0.975cm]$2u_2 {\cdot} (k_3{+}k_4)$}] {\midarrow} (c2.center);
	\draw[line width=0.5mm] (p1.center) -- (p4.center);
	\draw[line width=0.5mm] (p2.center) -- (p3.center);
\end{tikzpicture}
\caption{Parametrization of the loop momenta for the 2SF integral topology $\diagramnumberingsign40$ of table~\ref{tab: 4-loop results 2SF}, which in the even-parity sector depends on a different K3 surface than the integrals in sec.~\ref{sec:K31}.}
\label{fig: diag_K3_40}
\end{figure}
In particular, we can parametrize the integral following fig.~\ref{fig: diag_K3_40} with the integration order $\{ k_1, k_4, k_3, k_2 \}$, for which we introduce the 4 ISPs $z_{11}=k_2^2$, $z_{12}=(k_2-q)^2$, $z_{13}=2u_1 \cdot k_2$ and $z_{14}=2u_2 \cdot k_2$. Then, we find the leading singularity for the even-parity scalar integral to be
\begin{align}
& \LS \left( \begin{tikzpicture}[baseline={([yshift=-0.1cm]current bounding box.center)}] 
	\node[] (a) at (0,0) {};
	\node[] (a1) at (0.75,0) {};
	\node[] (a2) at (1.5,0) {};
	\node[] (c) at (0,-1) {};
	\node[] (c1) at (0.75,-1) {};
	\node[] (c2) at (1.5,-1) {};
	\node[] (p1) at ($(a)+(-0.2,0)$) {};
	\node[] (p2) at ($(c)+(-0.2,0)$) {};
	\node[] (p3) at ($(c2)+(0.2,0)$) {};
	\node[] (p4) at ($(a2)+(0.2,0)$) {};
	\draw[line width=0.15mm] (a.center) -- (c.center);
	\draw[line width=0.15mm] (a1.center) -- (c1.center);
	\draw[line width=0.15mm] (a2.center) -- (c2.center);
	\draw[line width=0.15mm] (a.center) -- (c2.center);
	\draw[line width=0.5mm] (p1.center) -- (p4.center);
	\draw[line width=0.5mm] (p2.center) -- (p3.center);
\end{tikzpicture} \right) \propto \LS \left( \int \frac{x^2 \, d z_{11} \, \cdots \, d z_{14}}{\sqrt{4z_{11}-z_{13}^2} \sqrt{4z_{11}-z_{14}^2} } \right. \nonumber \\
& \, \enspace \quad \times \left. \frac{1}{\sqrt{(x^2-1)^2 z_{12}^2 - 2 x (x^2+1) z_{12} z_{13} z_{14} + x^2 z_{13}^2 z_{14}^2} \, \sqrt{P_2(z_{11}, \dots, z_{14})}} \right),
\end{align}
where $P_2(z_{11}, \dots, z_{14})$ is a polynomial of overall degree 2, which is quadratic in all variables. Now, we can use the change of variables in eq.~\eqref{eq: change_of_variables_(z-r1)(z-r2)} to rationalize the first square root by changing from $z_{13}$ to $t_{13}$, as well as the second square root by changing from $z_{14}$ to $t_{14}$. Doing so, we obtain
\begin{align}
& \LS \left( \begin{tikzpicture}[baseline={([yshift=-0.1cm]current bounding box.center)}] 
	\node[] (a) at (0,0) {};
	\node[] (a1) at (0.75,0) {};
	\node[] (a2) at (1.5,0) {};
	\node[] (c) at (0,-1) {};
	\node[] (c1) at (0.75,-1) {};
	\node[] (c2) at (1.5,-1) {};
	\node[] (p1) at ($(a)+(-0.2,0)$) {};
	\node[] (p2) at ($(c)+(-0.2,0)$) {};
	\node[] (p3) at ($(c2)+(0.2,0)$) {};
	\node[] (p4) at ($(a2)+(0.2,0)$) {};
	\draw[line width=0.15mm] (a.center) -- (c.center);
	\draw[line width=0.15mm] (a1.center) -- (c1.center);
	\draw[line width=0.15mm] (a2.center) -- (c2.center);
	\draw[line width=0.15mm] (a.center) -- (c2.center);
	\draw[line width=0.5mm] (p1.center) -- (p4.center);
	\draw[line width=0.5mm] (p2.center) -- (p3.center);
\end{tikzpicture} \right) \propto \LS \left( \int \frac{x^2 \, t_{13} t_{14} \, d z_{11} d z_{12} d t_{13} d t_{14}}{\sqrt{P_7(z_{11}, z_{12}, t_{13}, t_{14})} \sqrt{P_{10}(z_{11}, z_{12}, t_{13}, t_{14})}} \right).
\end{align}
Here, $P_7(z_{11}, z_{12}, t_{13}, t_{14})$ and $P_{10}(z_{11}, z_{12}, t_{13}, t_{14})$ respectively denote degree-7 and degree-10 polynomials. Notably, the latter only depends quadratically on $z_{12}$. Therefore, we can use the transformation in eq.~\eqref{eq: change_of_variables_(z-r1)(z-r2)} with respect to $z_{12}$ to rationalize it, leading to
\begin{equation}
    \LS \left( \begin{tikzpicture}[baseline={([yshift=-0.1cm]current bounding box.center)}] 
	\node[] (a) at (0,0) {};
	\node[] (a1) at (0.75,0) {};
	\node[] (a2) at (1.5,0) {};
	\node[] (c) at (0,-1) {};
	\node[] (c1) at (0.75,-1) {};
	\node[] (c2) at (1.5,-1) {};
	\node[] (p1) at ($(a)+(-0.2,0)$) {};
	\node[] (p2) at ($(c)+(-0.2,0)$) {};
	\node[] (p3) at ($(c2)+(0.2,0)$) {};
	\node[] (p4) at ($(a2)+(0.2,0)$) {};
	\draw[line width=0.15mm] (a.center) -- (c.center);
	\draw[line width=0.15mm] (a1.center) -- (c1.center);
	\draw[line width=0.15mm] (a2.center) -- (c2.center);
	\draw[line width=0.15mm] (a.center) -- (c2.center);
	\draw[line width=0.5mm] (p1.center) -- (p4.center);
	\draw[line width=0.5mm] (p2.center) -- (p3.center);
\end{tikzpicture} \right) \propto \LS \left( \int \frac{x^2 \, dz_{11} dt_{12} dt_{13} dt_{14}}{\sqrt{P_{14}(z_{11},t_{12},t_{13},t_{14})}} \right).
\end{equation}
Even though $P_{14}(z_{11},t_{12},t_{13},t_{14})$ defines a degree-14 polynomial, it is nonetheless quadratic in $z_{11}$. Rationalizing it with eq.~\eqref{eq: change_of_variables_(z-r1)(z-r2)} reveals a simple pole at $t_{11}=0$. Taking the residue yields once again a single square root over a degree-6 polynomial, but which is only quadratic in $t_{12}$. Repeating the change of variables and taking the residue at $t'_{12}=0$, we finally obtain a degree-8 polynomial in $t_{13}$ and $t_{14}$. Rescaling $t_{14} \to t_{14}/t_{13}$, and changing $\{ t_{13}, t_{14} \} \to \{ \sqrt{t_1}, t_2 \}$, we finally arrive at
\begin{equation}
    \LS \left( \begin{tikzpicture}[baseline={([yshift=-0.1cm]current bounding box.center)}] 
	\node[] (a) at (0,0) {};
	\node[] (a1) at (0.75,0) {};
	\node[] (a2) at (1.5,0) {};
	\node[] (c) at (0,-1) {};
	\node[] (c1) at (0.75,-1) {};
	\node[] (c2) at (1.5,-1) {};
	\node[] (p1) at ($(a)+(-0.2,0)$) {};
	\node[] (p2) at ($(c)+(-0.2,0)$) {};
	\node[] (p3) at ($(c2)+(0.2,0)$) {};
	\node[] (p4) at ($(a2)+(0.2,0)$) {};
	\draw[line width=0.15mm] (a.center) -- (c.center);
	\draw[line width=0.15mm] (a1.center) -- (c1.center);
	\draw[line width=0.15mm] (a2.center) -- (c2.center);
	\draw[line width=0.15mm] (a.center) -- (c2.center);
	\draw[line width=0.5mm] (p1.center) -- (p4.center);
	\draw[line width=0.5mm] (p2.center) -- (p3.center);
\end{tikzpicture} \right) \propto \frac{x^2}{x^2-1} \, \LS \! \left( \int \! \frac{dt_1 dt_2}{\sqrt{(x(1+t_1)(t_1+t_2^2){-}(1+x^2)t_1t_2)^2-4x^2t_1^2t_2^2}} \right)\!.
\label{eq: Q6}
\end{equation}
We observe the appearance of a polynomial of degree 6 in two variables which, based on the analysis in sec.~\ref{sec:review_DE_LS}, defines a period integral over a K3 surface.
However, it corresponds to a different K3 surface than the 4PM geometry which made an appearance in sec.~\ref{sec:K31} for multiple 5PM 1SF integral topologies; see refs.~\cite{Klemm:2024wtd,Duhr:2025lbz} for details. Thus, compared to the 1SF order, at the 2SF order we not only observe the appearance of a new three-dimensional Calabi--Yau geometry, but also the 1SF K3 surface as well as a second K3 surface.

\section{Conclusions}
\label{sec:conclusions}

In this paper, we have classified all geometries that appear in Feynman integrals relevant for $2 \rightarrow 2$ black-hole scattering at the fifth post-Minkowskian order (5PM), to all orders in the self-force (SF) expansion. These geometries determine which special functions can appear in the gravitational-wave observables, both for the conservative and the dissipative dynamics. The classification of geometries was achieved by a substantial reduction of the Feynman integral topologies that need to be analyzed, from 16,596 to only 70 at four loops, followed by a calculation of their leading singularities. Aside from integrals of polylogarithmic type, we encounter four distinct, non-trivial geometries: two three-dimensional Calabi--Yau geometries, and two K3 surfaces. They occur in 8 independent Feynman integral topologies, which are depicted in fig.~\ref{fig: non-trivial_diagrams_4loop}.

At 5PM 1SF order, the full amplitude has recently been completed~\cite{Driesse:2024xad,Bern:2024adl,Driesse:2024feo}, including a three-dimensional Calabi--Yau geometry in the dissipative sector and a K3 surface -- previously encountered at 4PM order~\cite{Ruf:2021egk,Dlapa:2022wdu,Frellesvig:2024zph} -- in both the conservative and dissipative regimes. Our findings from the classification of non-trivial geometries are in complete agreement with these results. 

At 5PM 2SF order, our classification narrowly restricts which geometries can occur. It predicts the reappearance of the 4PM K3 surface in the dissipative sector, along with a different K3 surface and a different three-dimensional Calabi--Yau geometry in the conservative regime.
Recently, $\varepsilon$-factorized differential equations for the sectors involving these additional geometries have been constructed~\cite{Frellesvig:2024rea,Duhr:2025lbz}. Combined with the complete classification presented in this work, these advances remove the last remaining obstacles from non-trivial geometries for a full 5PM 2SF computation.

Moreover, we provide the leading singularities of the relevant Feynman integrals at 2SF order in table~\ref{tab: 4-loop results 2SF}. These should facilitate finding an $\varepsilon$-factorized differential equation for the remaining polylogarithmic integrals and accelerate the full amplitude calculation, since the leading singularity can be used to construct a basis of pure master integrals~\cite{Henn:2014qga, Primo:2016ebd, WasserMSc, Dlapa:2021qsl}. We note, however, that the occurrence of non-planar integral topologies makes the full 2SF computation particularly challenging.

One of the advantages of the method presented in this paper, which is based on the Baikov representation and leading singularities, is that it is computationally light. Alternatively, one can perform the IBP reductions and compute the differential equations in order to obtain the Picard--Fuchs operator annihilating a given Feynman integral, which also characterizes its underlying geometry. However, IBP reductions already became a major bottleneck for the full calculations of observables at 5PM 1SF order~\cite{Driesse:2024xad,Bern:2024adl,Driesse:2024feo} and are expected to be computationally more expensive at the 2SF order. It would thus be highly desirable to avoid this step via a bootstrap approach that leverages the understanding of the occurring geometries and corresponding special functions. We leave this line of research for future work. 

Finally, it would be very interesting to extend our analysis to 6PM order, i.e.\ to five loops, in view of the upcoming third generation of gravitational-wave detectors~\cite{Buonanno:2022pgc}, as well as for QCD amplitudes for Standard Model particle phenomenology~\cite{Bargiela:toapp}.

\section*{Acknowledgements}

We thank Piotr Bargiela, Zvi Bern, Christoph Dlapa, Enrico Herrmann, Gustav Jakobsen, Zhengwen Liu, Andres Luna, Robin Marzucca, Andrew McLeod, Jan Plefka, Sebastian Pögel, Rafael Porto, Michael Ruf, Cristian Vergu, Matt von Hippel, Stefan Weinzierl and Tong-Zhi Yang for fruitful discussions. 

The work of HF, RM and MW was supported by the research grant 00025445 from Villum Fonden. HF has received funding from the European Union's Horizon 2020 research and innovation program under the Marie Sk{\l}odowska-Curie grant agreement No. 847523 `INTERACTIONS'. HF was moreover supported in part by the National Natural Science Foundation of China. MW was further supported by the Sapere Aude: DFF-Starting Grant 4251-00029B.

\bibliography{References}
\bibliographystyle{JHEP}

\end{document}